\documentclass[fleqn,10pt]{wlscirep}
\usepackage[utf8]{inputenc}
\usepackage[T1]{fontenc}
\usepackage{color}
\definecolor{red}{rgb}{1,0,0}

\title{Small-worldness favours network inference}

\author[1,*]{Rodrigo A. Garc{\'i}a}
\author[1]{Arturo C. Martí}
\author[1]{Cecilia Cabeza}
\author[1]{{Nicol{\'a}s Rubido}}
\affil[1]{Universidad de la Rep{\'u}blica, Instituto de F{\'i}sica de Facultad de Ciencias, Montevideo, 11400, Uruguay.}

\affil[*]{rgarcia@fisica.edu.uy}


\begin{abstract}
A main goal in the analysis of a complex system is to infer its underlying network structure from time-series observations of its behaviour. The inference process is often done by using bi-variate similarity measures, such as the cross-correlation (CC), however, the main factors favouring or hindering its success are still puzzling. Here, we use synthetic neuron models in order to reveal the main topological properties that frustrate or facilitate inferring the underlying network from CC measurements. Specifically, we use pulse-coupled Izhikevich neurons connected as in the Caenorhabditis elegans neural networks as well as in networks with similar randomness and small-worldness. We analyse the effectiveness and robustness of the inference process under different observations and collective dynamics, contrasting the results obtained from using membrane potentials and inter-spike interval time-series. We find that overall, small-worldness favours network inference and degree heterogeneity hinders it. In particular, success rates in C. elegans networks -- that combine small-world properties with degree heterogeneity -- are closer to success rates in Erd{\"o}s-R{\'e}nyi network models rather than those in Watts-Strogatz network models. These results are relevant to understand better the relationship between topological properties and function in different neural networks.

\end{abstract}
\begin{document}

\flushbottom
\maketitle

\thispagestyle{empty}

\section*{Introduction}\label{sec_intro}
Network Neuroscience seeks to unravel the complex relationship between functional connectivity in neural systems (i.e., the correlated neural activity) and their underlying structure (e.g., the brain's connectome); among other goals \cite{bassett2017network,medaglia2015cognitive,sporns2014contributions,bullmore2009complex,sporns2005human}. The functional connectivity is responsible for many tasks, such as segregation, transmission, and integration of information \cite{deco2015rethinking,sporns2013network}. These (and other) tasks are found to be optimally performed by neural structures that show small-world properties \cite{bullmore2009complex,medaglia2015cognitive,deco2015rethinking}, which range from human brain connectomics \cite{sporns2005human,jorgenson2015brain,haimovici2013brain} to the Caenorhabditis elegans (C. elegans) nematode neural networks \cite{varier2011neural,ren2010stdp,antonopoulos2016dynamical}. In general, structural networks have been revealed by tracing individual neural processes, e.g., by diffusion tensor imaging or electron-microscopy -- methods that are typically unfeasible for large neural networks. On the other hand, functional networks are revealed by performing reverse engineering on the time-series measurements of the neural activity, e.g., by using EEG or EMG recordings. These methods are known as network inference and are mainly affected by data availability and precision.

In general, network inference has been approached by means of bi-variate similarity measures, such as the pair-wise Cross-Correlation \cite{eguiluz2005scale,perrin2012electroconvulsive,haimovici2013brain}, Granger Causality \cite{bressler2011wiener,ge2012characterizing,sommerlade2012inference}, Transfer Entropy \cite{sun2015causal,villaverde2014mider,tung2007inferring}, and Mutual Information \cite{basso2005reverse,rubido2014exact,tirabassi2015inferring,bianco2016successful}, to name a few. The main idea behind the similarity approach is that, units sharing a direct connection (namely, a functional or structural link exists that joins them) have particularly similar dynamics, whereas units that are indirectly connected (namely, a functional or structural link joining them is absent) are less likely to show similar dynamics. Although intuitive, this approach has found major challenges in neural systems due to their complex behaviour and structure connectedness, resulting in highly correlated dynamics from indirectly connected units and loosely correlated dynamics for directly connected units. Moreover, because most works have focused on maximising the inference success (in relation to its ability to discover the structural network) and/or optimising its applicability \cite{kramer2009network,friedman2000using,rubido2014exact,bianco2016successful}, we are still unaware of which are the main underlying mechanisms that affect the inference results. Namely, differentiating the underlying structure with the functional connectivity -- particularly with respect to establishing which of the different network properties are mainly responsible for hindering inference success rates.

In this work, we reveal that the degree of small-worldness is directly related to the success of correctly inferring the network of synthetic neural systems when using bi-variate similarity analysis. Our neural systems are composed of pulse-coupled Izhikevich maps and connected in network ensembles with different small-worldness levels -- but statistically similar to the C. elegans neural networks. The inference process is done by means of the pair-wise cross-correlation between the neurons' activity. We assess the inference effectiveness by means of receiver operating characteristic (ROC) analysis and, in particular, the true positive rate ($TPR$), which we show is the only relevant quantity under our inference framework. Our findings show that the $TPR$ peaks around a critical coupling strength where the system transitions from synchronous bursting dynamics to a spiking incoherent regime. Specifically, we find that the highest $TPR$ is for networks with significant small-worldness level. We analyse these results in terms of different topology choices, collective dynamics, neural activity observations (inter-spike intervals or membrane potentials), and time-series length. We expect that these results will help to understand better the role of small-worldness in brain networks, but also in other complex systems, such as climate networks \cite{tsonis2006networks,donges2009complex,donges2009backbone}.


\section*{Results} \label{sec_results}
We infer the underlying network of a synthetic neural system by creating a binary matrix of $1$s and $0$s from the pair-wise cross-correlation (CC) matrix of the signal-measurements. The resultant binary matrix represents the inferred connections that the neurons composing the system share, which we obtain by applying a threshold to the CC matrix. The threshold assumes that a strong [weak] similarity in the measured signals, i.e., a CC value above [below] the threshold, correspond to a $1$ [$0$] in the inferred adjacency matrix, suggesting that a direct [indirect] structural connection exists. In spite of this (seemingly) over-simplification, this binary process is broadly used in network inference \cite{sporns2005human,bullmore2009complex,haimovici2013brain,rubido2014exact,bianco2016successful} and it tends to keep the most relevant information from the underlying connectivity. Moreover, when the underlying network is known, it allows to quantify how poorly or efficiently the bi-variate method performs in terms of the receiving operation characteristic (ROC) analysis \cite{brown2006receiver,fawcett2006introduction,rogers2005bayesian}. In particular, we set the threshold such that the inferred network has the same density of connections, $\rho = 2M/N\,(N - 1)$, as the underlying structure, where $N$ is the network size and $M$ is the number of existing links. This means that we assume an a priori (minimal) knowledge about the underlying structure, namely, we require knowing $\rho$ in order to choose the threshold.

The true positive rate, also known as sensitivity, is the proportion of correctly identified connections with respect to the total of existing connections \cite{brown2006receiver}, i.e., $TPR \equiv TP/\left( TP + FN \right)$, where $TP$ is the number of true positives and $FN$ is the number of false negatives. This quantity is part of the ROC analysis, which includes the true negative rate, $TNR$, false positive rate, $FPR$, and false negative rate, $FNR = 1 - TPR$. Taken together, these variables quantify the performance of any method. However, when fixing the inferred network's density of connections, $\rho$, to match that of the underlying network, we can show that the $TPR$ is the only relevant variable in the ROC analysis -- all remaining quantities can be expressed in terms of the $TPR$ and $\rho$. For example, the $TNR = 1 - FPR$, also known as specificity, can be expressed in terms of the $TPR$ and $\rho$ by
\begin{equation}
 TNR \equiv \frac{TN}{TN + FP} = \frac{TP}{TN + FP} + \frac{N\,(N - 1)/2 - 2M}{N\,(N - 1)/2 - M} = \left(\frac{\rho}{1 - \rho}\right) TPR + \left(\frac{1 - 2\rho}{1 - \rho}\right) = 1 - FPR,
    \label{eq_TNR}
\end{equation}
where we use the fact that $TP + FN = M$ is the number of existing connections, $TN + FP = N\,(N - 1)/2 - M$ is the number of non-existing connections, $\rho = 2M/N\,(N - 1)$ is the density of connections, and $TP + FP = M$ is the number of connections we keep fixed for the inferred matrix in order to maintain $\rho$ invariant. Hence, as a result of our threshold choice, $FN = FP$, implying that the $FNR \equiv FN/\left( FN + TP \right)$ is identical to the false discovery rate, $FDR \equiv FP/\left( FP + TP \right)$, and that the precision, $PPV \equiv TP/\left( TP + FP \right)$ is identical to the $TPR$. Overall, these relationships mean that in order to quantify the inference success or failure, we can solely focus on studying how the $TPR$ changes as the dynamical parameters and network structure change.

The following results are derived from time-series measurements of pulse-coupled Izhikevich maps interacting according to different network structures and coupling strengths, where each map's uncoupled dynamic is set to bursting (see \href{sec_methods}{Methods} for details on the map and network parameters). Pulse coupling is chosen because of its generality, which has been shown to allow the representation of several biophysical interactions \cite{izhikevich2003simple,izhikevich2004model,ibarz2011map}, and single parameter tuning, i.e., the coupling strength, $\epsilon$. In particular, we register the neurons' membrane potentials (signals coming from the electrical impulses) and inter-spike intervals (time windows between the electrical pulses) of $10$ randomly-set initial conditions and $T = 7\times10^4$ iterations, from which we discard the first $2\times10^4$ iterations as transient (we also analyse the effects of keeping shorter time-series).

Without losing generality, we restrict our analyses to connecting the neurons (maps) in symmetric Erd{\"o}s-R{\'e}nyi (ER) \cite{erdos1960evolution} and Watts-Strogatz (WS) \cite{watts1998collective} network ensembles that have identical size, $N$, and sparse density of connections, $\rho$, to that of the C. elegans frontal and global neural network \cite{varier2011neural,ren2010stdp,antonopoulos2016dynamical}. Specifically, when constructing the ensembles we set $N_f = 131$ with $\rho_f \simeq 0.08$ (which corresponds to setting a mean degree, $\left\langle \overline{k} \right\rangle_f \simeq 10.5$), or $N_g = 277$ with $\rho_g \simeq 0.05$ (which corresponds to setting a mean degree, $\left\langle \overline{k} \right\rangle_g \simeq 13.8$), respectively. The reason behind this choice is that the C. elegans neural networks are one of the most cited examples of real-world small-world networks \cite{watts1998collective,antonopoulos2016dynamical,sporns2005human,varier2011neural,ren2010stdp}, showing small average shortest paths connecting nodes and high clustering. Namely, the C. elegans neural networks have a high small-worldness coefficient $\sigma$, defined as the normalised ratio between the clustering coefficient and average path length \cite{humphries2008network,telesford2011ubiquity}, but also show an heterogeneous degree distribution. More importantly, these network ensembles constitute a controlled setting where to compare and distinguish the main topological factors favouring or hindering the inference success, providing us with a reproducible framework to modify the network properties within each realisation.

\begin{figure}[htbp]
 \begin{center}
  \begin{minipage}{0.49\textwidth}
   \begin{center}
    {\bf (a)}\\ \includegraphics[scale= 0.55]{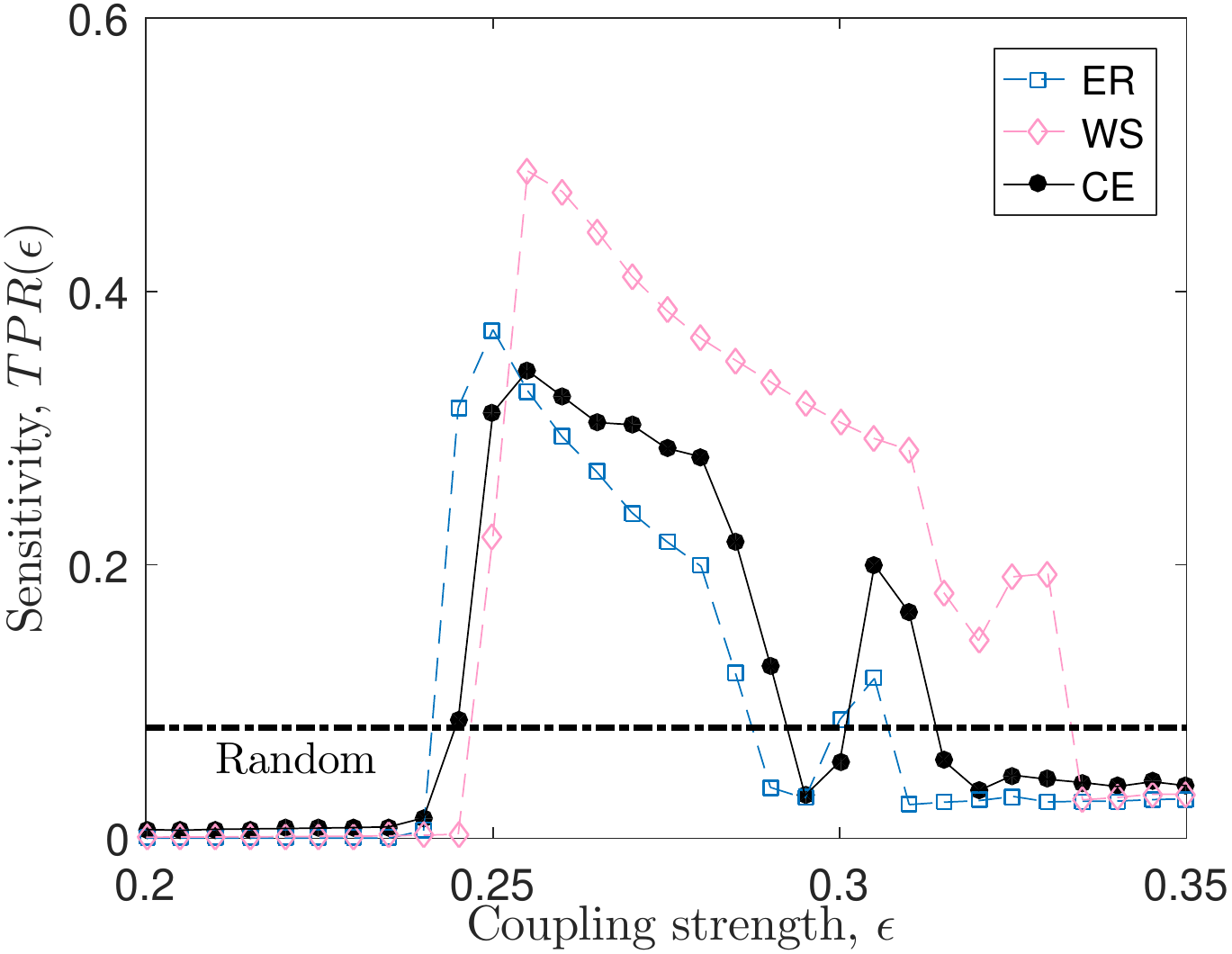}
   \end{center}
  \end{minipage} \hspace{0.2pc}
  \begin{minipage}{0.49\textwidth}
   \begin{center}
    {\bf (b)}\\ \includegraphics[scale= 0.55]{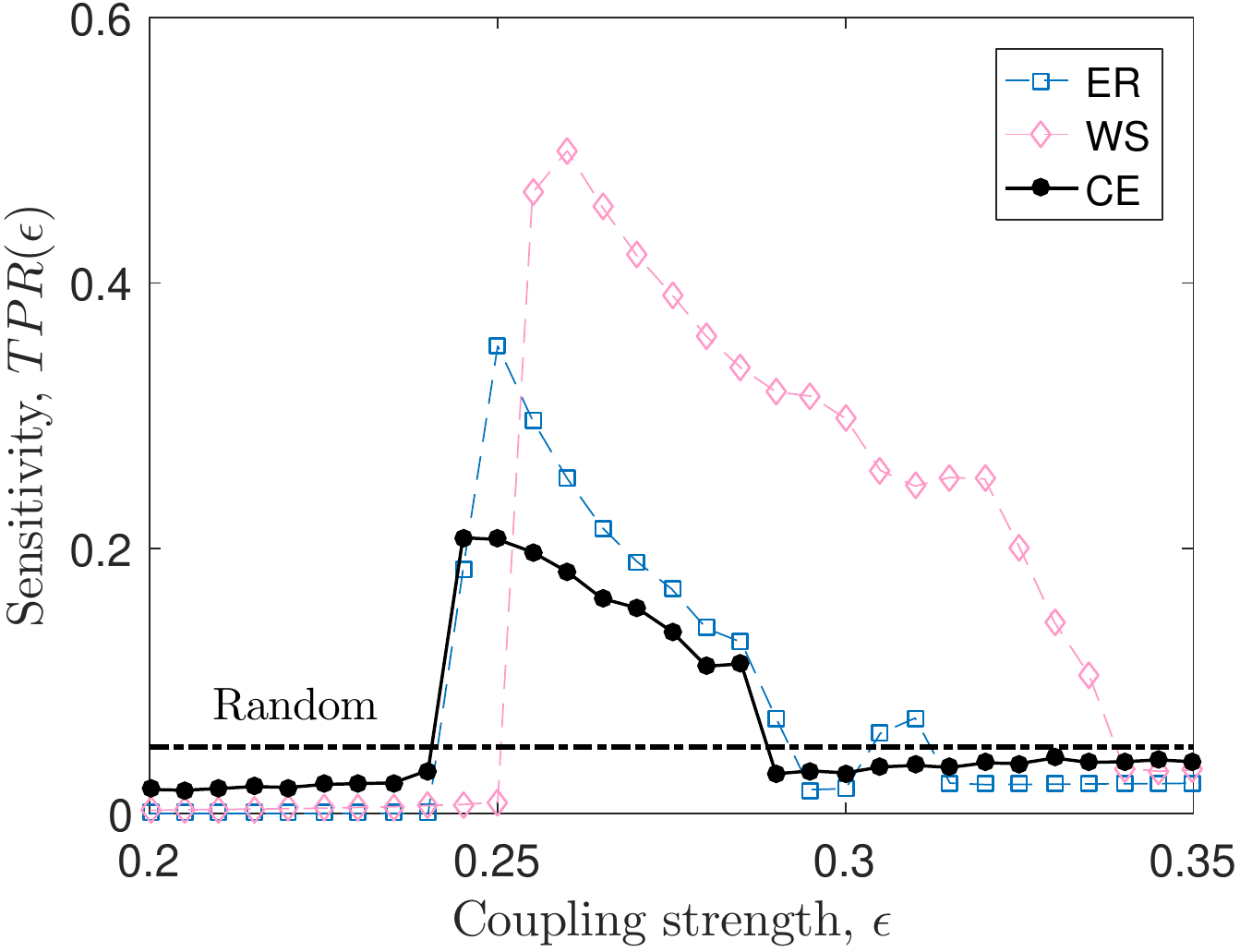}
   \end{center}
  \end{minipage}
 \end{center}
    \caption{{\bf Network inference success rates for different networks, coupling strengths, and sizes}. Panel {\bf (a)} [Panel {\bf (b)}] shows the true positive rate, $TPR$, as a function of the coupling strength, $\epsilon$, between $N = 131$ [$N = 277$] pulse-coupled Izhikevich maps connected in Erd{\"o}s-R{\'e}nyi (ER), Watts-Strogatz (WS), or C. elegans (CE) frontal [global] neural networks; map parameters are set such that the isolated dynamics is bursting (see \href{sec_methods}{Methods}). The $TPR$ values for the ER and WS come from averaging the results over $10$ initial conditions and $20$ network realisations with similar topological properties to that of the CE (i.e., number of nodes, average degree, and density of connections). For the CE, the results are averaged only on the initial conditions. The $TPR$ is found by comparing the true underlying network with the binary matrix obtained from cross-correlating the membrane potential time-series ($T = 5\times10^4$ iterations) and fixing a threshold such that the inferred density of connections $\rho_f$ matches that of the CE, with $\rho_f \simeq 0.08$ in panel {\bf (a)} and $\rho_f \simeq 0.05$ in panel {\bf (b)}. The horizontal dashed line in both panels is the random inference $TPR$, namely, the null hypothesis.}
 \label{FIG:TPR_vs_Ep_CE_WS_ER}
\end{figure}

We find that the resultant average sensitivity from these network ensembles is more significant, robust, and reliable on WS ensembles than on ER ensembles, pointing to a fundamental importance of the underlying small-worldness for a successful inference. In particular, Fig.~\ref{FIG:TPR_vs_Ep_CE_WS_ER} shows the resultant success rates for ER (dotted lines with unfilled squares) and WS (dotted lines with unfilled diamonds) ensembles, plus, a comparison with the results we obtain when using the C. elegans (CE) neural frontal (left panel) and global (right panel) network structure (continuous lines with filled circles). Specifically, Fig.~\ref{FIG:TPR_vs_Ep_CE_WS_ER}{\bf (a)} and {\bf (b)} show the ensemble-averaged $TPR$ results for $N = 131$ and $N = 277$ pulse-coupled Izhikevich maps, respectively, as a function of the coupling strength, $\epsilon$, between the maps. From both panels we also note that the CE overall resultant success rates are closer to the ER ensemble-averaged $TPR$ results than to the WS ensemble-averaged $TPR$ results -- in spite of the CE small-worldness coefficient for the $N=131$ networks being the same as the WS, $\sigma=2.80$. The results in Fig.~\ref{FIG:TPR_vs_Ep_CE_WS_ER} show how important the underlying degree distribution and small-worldness are in the generation of collective dynamics that can be analysed by means of a bi-variate inference method with a sufficiently high success rate. 

\begin{figure}[htbp]
 \begin{center}
  \begin{minipage}{0.49\textwidth}
   \begin{center}
    {\bf (a)}\\ \includegraphics [scale= 0.55] {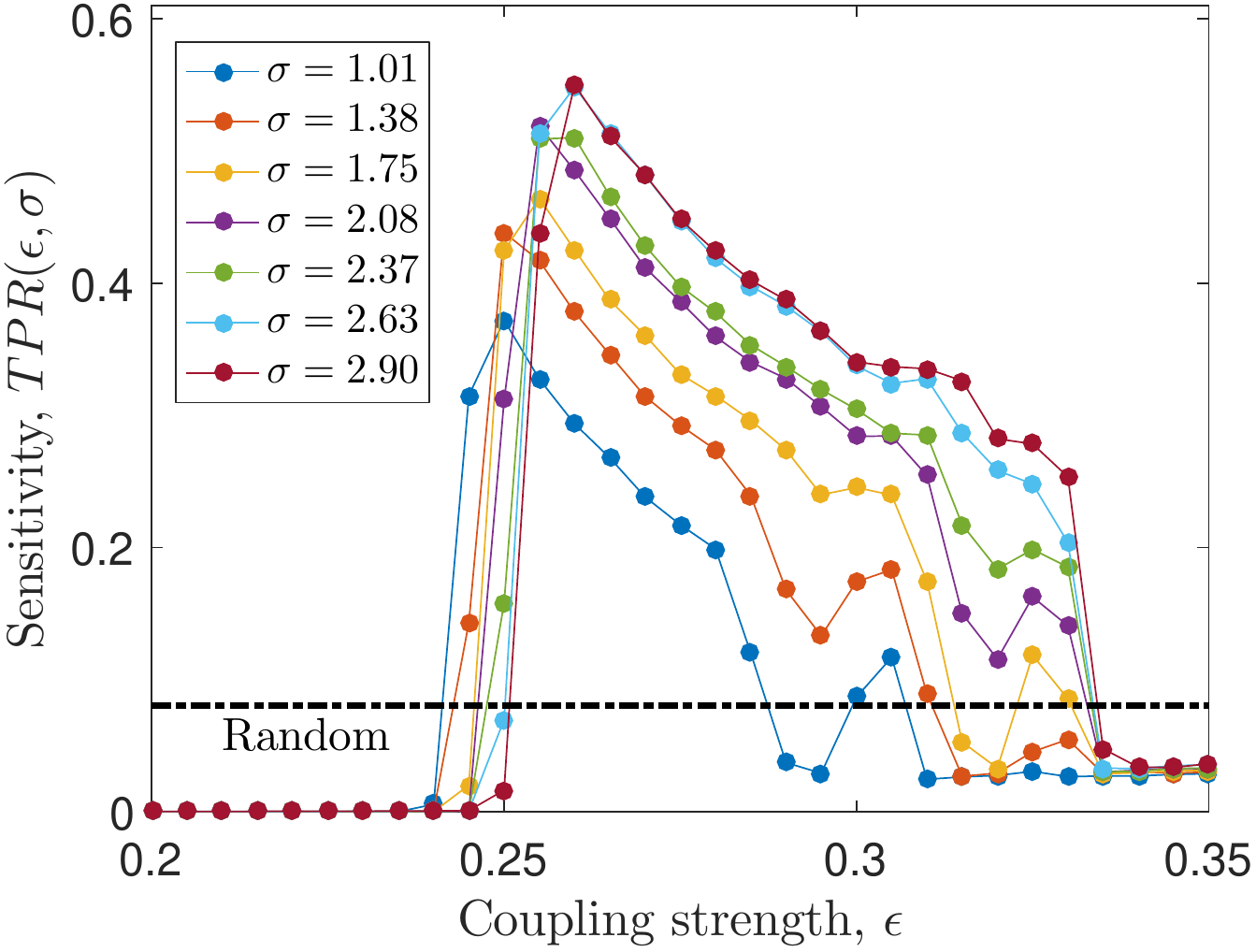}
   \end{center}
  \end{minipage} \hspace{0.2pc}
  \begin{minipage}{0.49\textwidth}
   \begin{center}
    {\bf (b)}\\ \includegraphics [scale= 0.55] {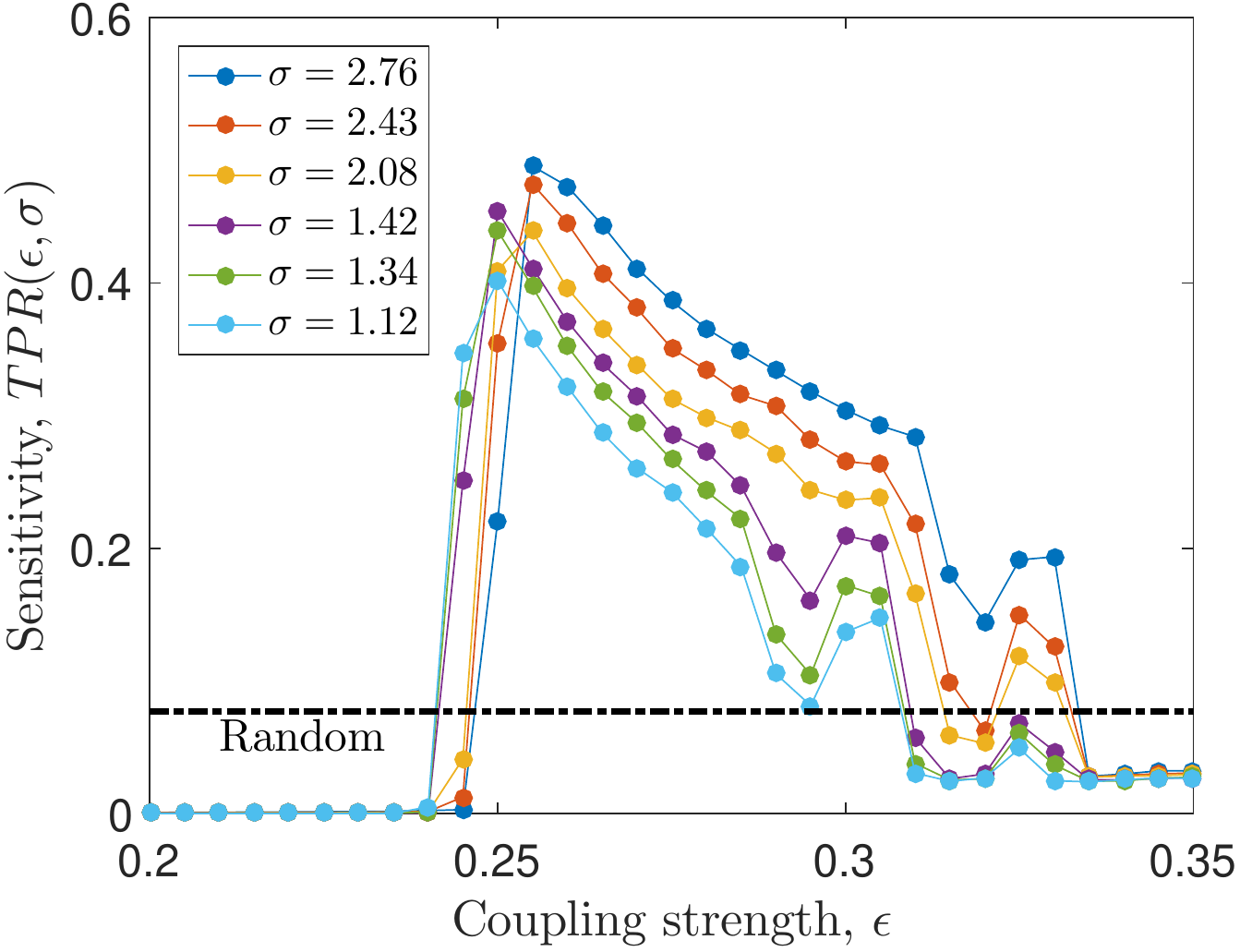}
   \end{center}
  \end{minipage}
 \end{center}
    \caption{ {\bf Network inference success rates as a function of coupling strength and small-worldness coefficient}. Using map and network parameters set as in Fig.~\ref{FIG:TPR_vs_Ep_CE_WS_ER}, panels {\bf (a)} and {\bf (b)} show the ensemble and initial-condition averaged $TPR$ as function of $\epsilon$ for $N = 131$ pulse-coupled Izhikevich maps in Erd{\"o}s-R{\'e}nyi (ER) and Watts-Strogatz (WS) network ensembles, respectively. A successive rewiring process \cite{maslov2004detection} is done to each network realisation in order to change its small-worldness coefficient, $\sigma$, whilst maintaining the underlying density of connections and degree distribution invariant. The colour code indicates the resultant $\sigma$ for each rewiring step that increases [panel {\bf (a)}] or decreases [panel {\bf (b)}] the network small-worldness. }
 \label{FIG:TPR_vs_Sg}
\end{figure}

In order to critically explore the significance that the underlying small-worldness has on the resultant inference, we fix the degree distribution and density of connections as we increase [decrease] $\sigma$ in each of the $20$ underlying ER [WS] network realisations using the rewiring method proposed in Ref.\cite{maslov2004detection}. Figure~\ref{FIG:TPR_vs_Sg} shows the resultant network inference -- quantified by the $TPR(\epsilon,\sigma)$ -- after we make the isolated changes in the small-worldness coefficient, $\sigma$, of the underlying structure for the $N = 131$ pulse-coupled Izhikevich maps (similar results are found for $N = 277$). The ensemble-averaged inference results (colour coded curves) that we get from making this topological change to $\sigma$ on the underlying ER and WS networked system are shown in Fig.~\ref{FIG:TPR_vs_Sg}{\bf (a)} and {\bf (b)}, respectively. We can see from these panels that the highest $TPR$ values are achieved for the largest $\sigma$ values, meaning that the best inference happens for networks with large $\sigma$. Also, we can see that there is a broad coupling strength interval ($0.25 \lesssim \epsilon \lesssim 0.33$) for both network classes that allows us to infer better than making blind random inference (dashed horizontal lines). From these panels, we note that network inference effectiveness increases robustly (namely, regardless of parameter changes) and significantly (namely, reliably across ensembles and random initial conditions) as the small-worldness, $\sigma$, of the underlying structure is increased -- whilst keeping its density of connections and degree distribution invariant. Consequently, in order to increase the inference success rates in the sparse ER networks, we need to increase the local clustering inter-connecting the maps. On the contrary, WS networks show optimal inference efficiency without modifying their clustering because of their inherent large small-worldness coefficient.

\begin{figure}[htbp]

 \begin{center}
  \begin{minipage}{0.32\textwidth}
   \begin{center}
    {\bf (a)}\\ \includegraphics [width= 5.78 cm] {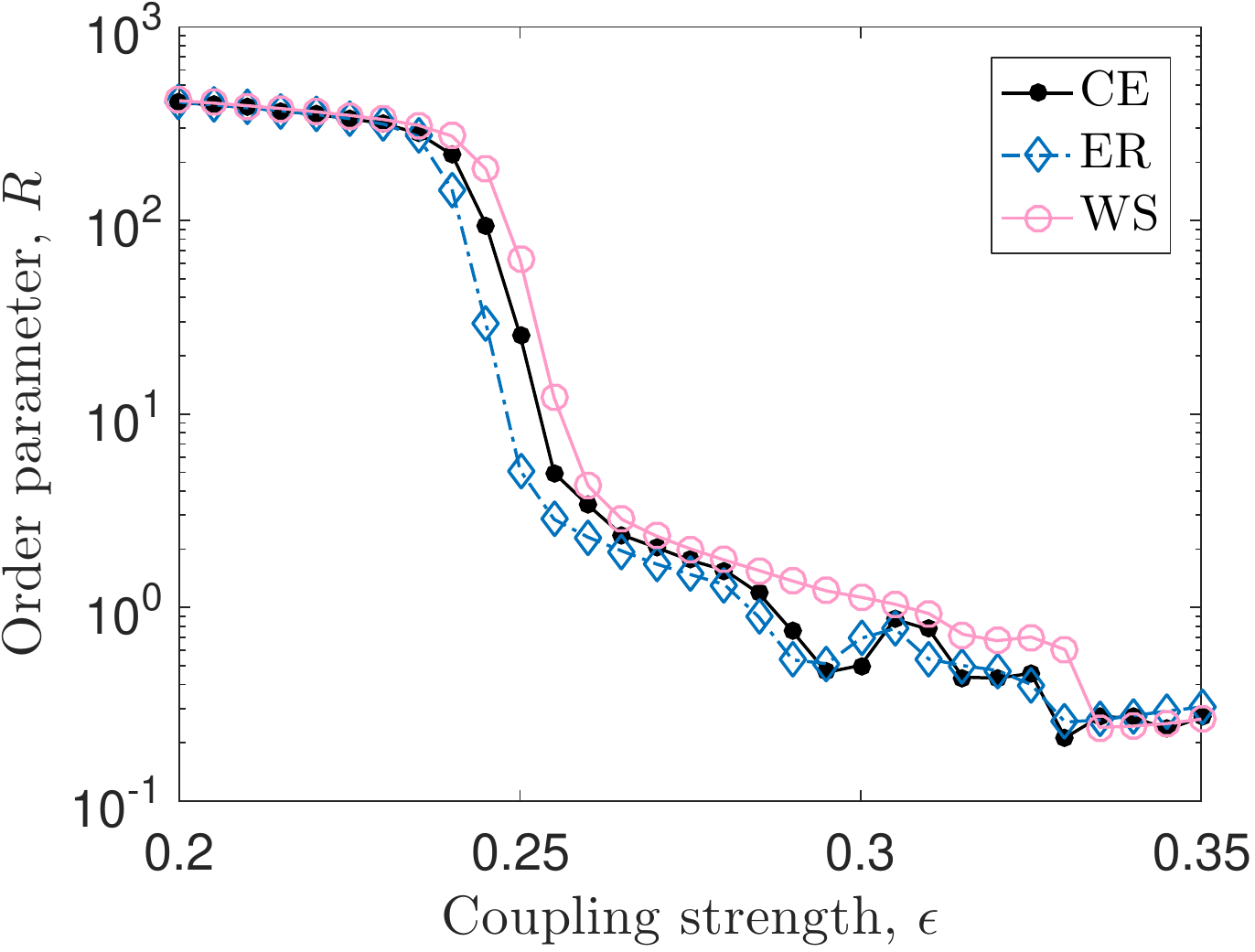} \\ 
    {\bf(d)} 
    \\  \includegraphics [width= 5.8 cm] {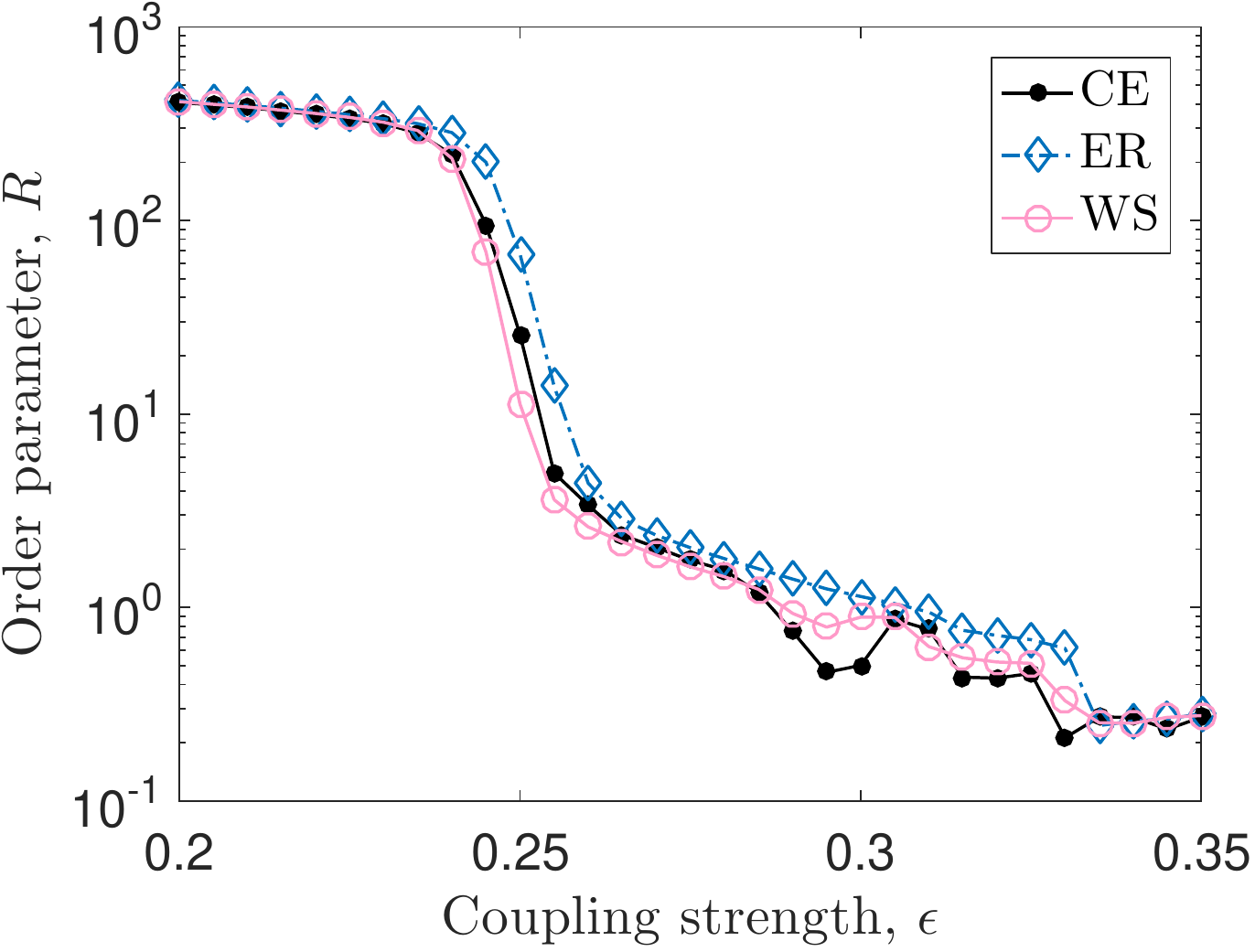}
   \end{center}
  \end{minipage} \hspace{0.2pc}
  \begin{minipage}{0.32\textwidth}
   \begin{center}
    {\bf (b)}\\ \includegraphics [width= 5.8 cm] {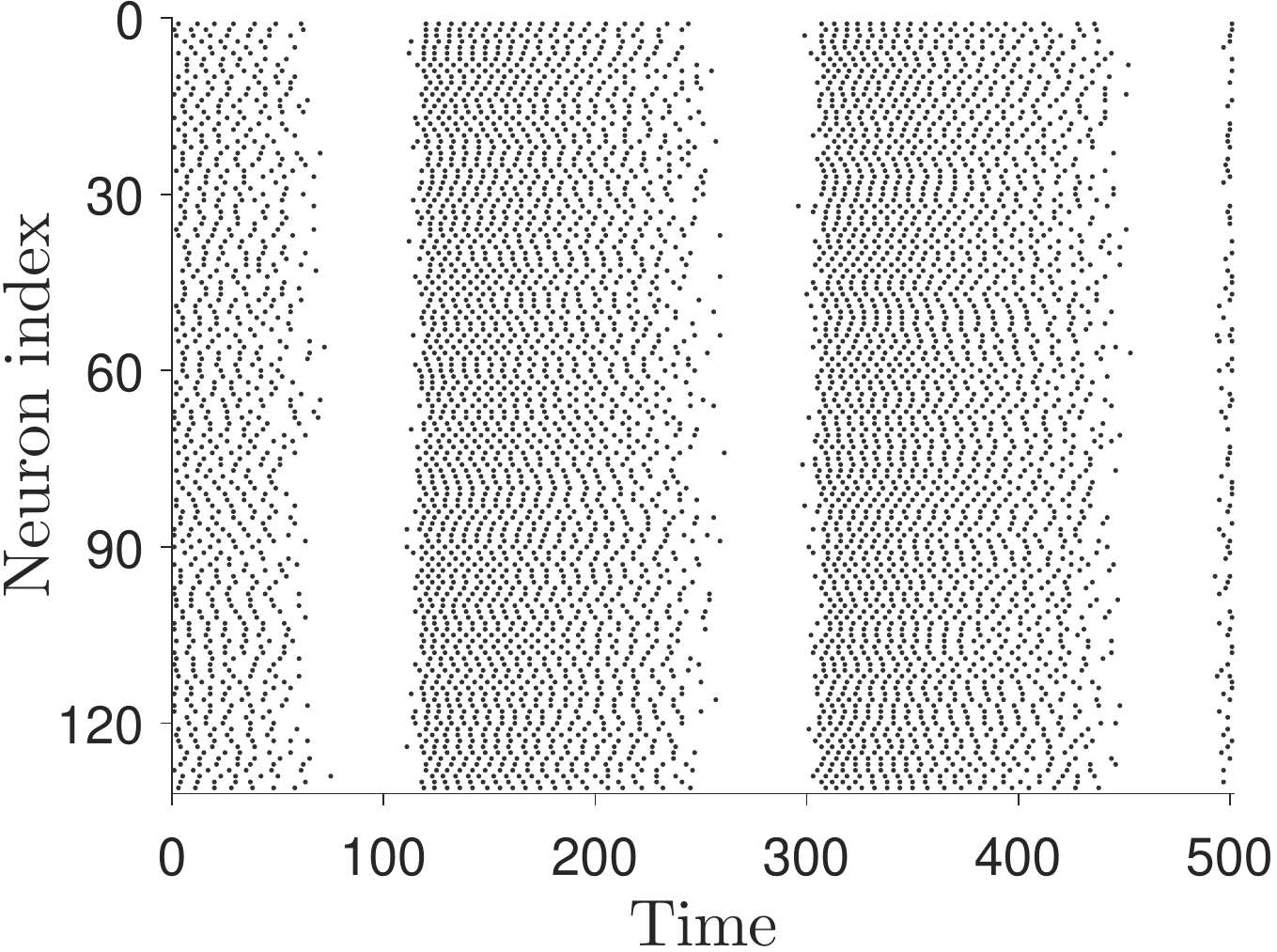} \\ {\bf(e)} \\  \includegraphics [width= 5.8 cm] {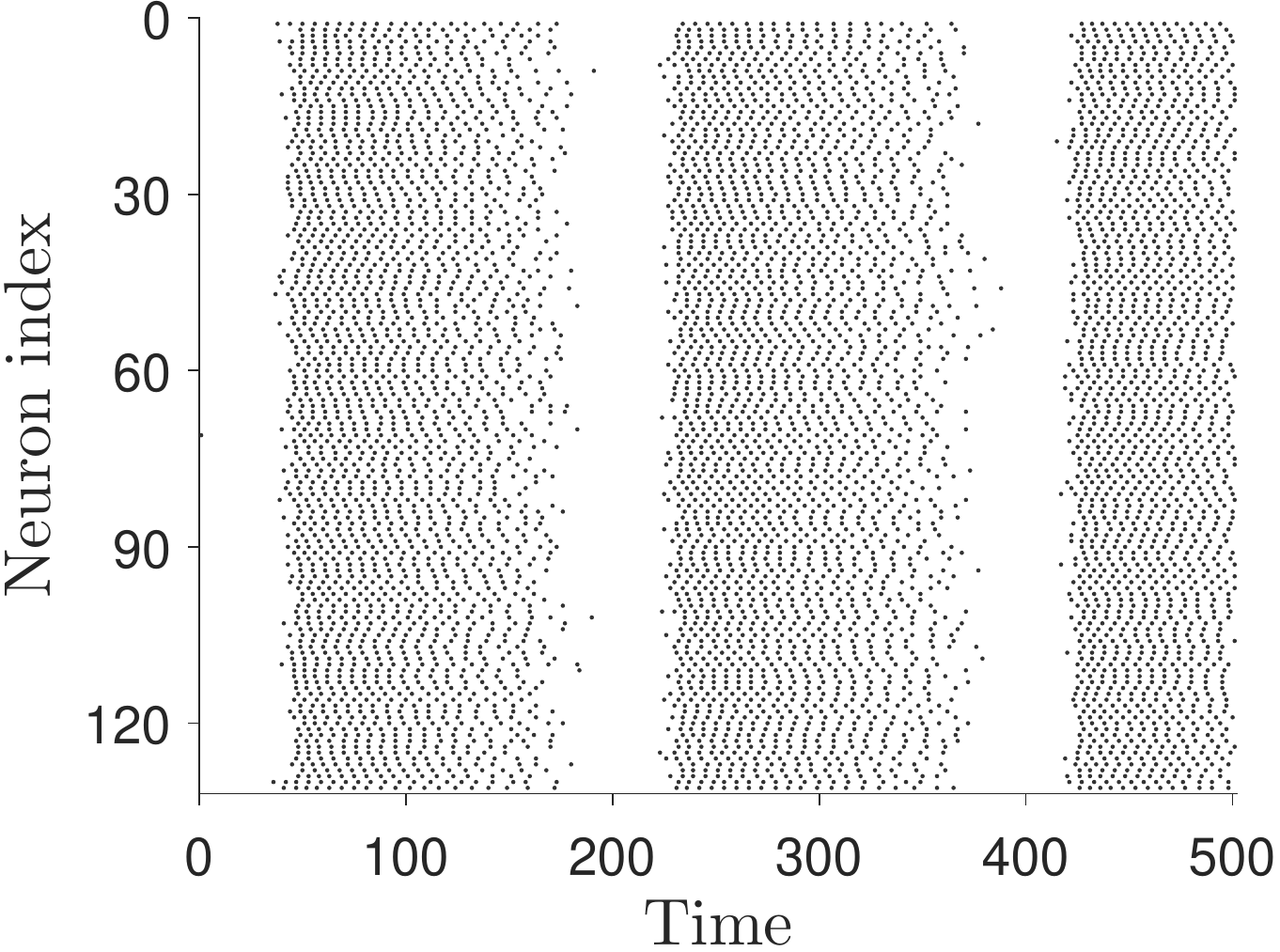}
   \end{center}
  \end{minipage}  \hspace{0.2pc}
  \begin{minipage}{0.32\textwidth}
   \begin{center}
    {\bf (c)}\\ \includegraphics [width= 5.8 cm] {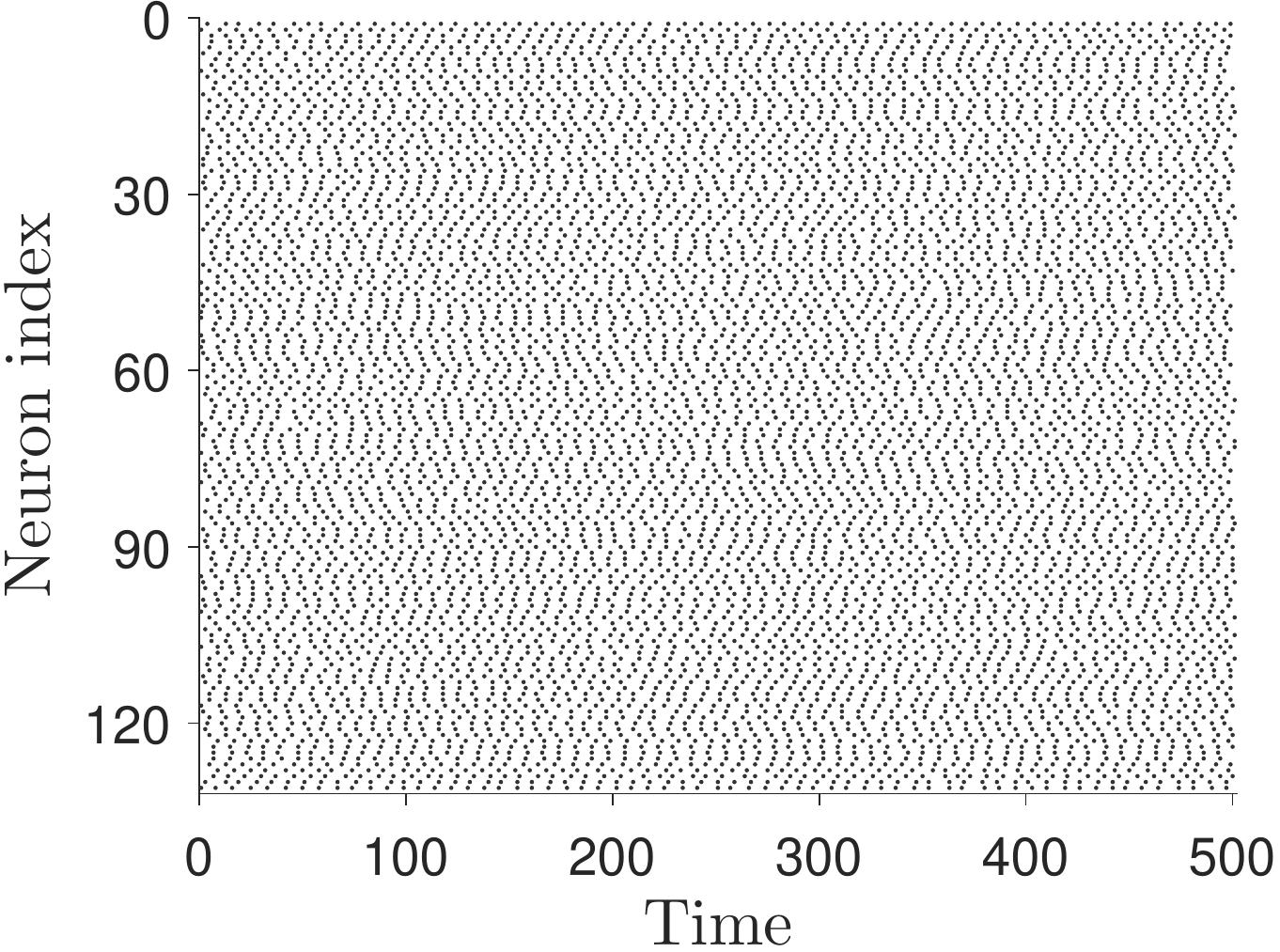} \\ {\bf(f)} \\  \includegraphics [width= 5.8 cm] {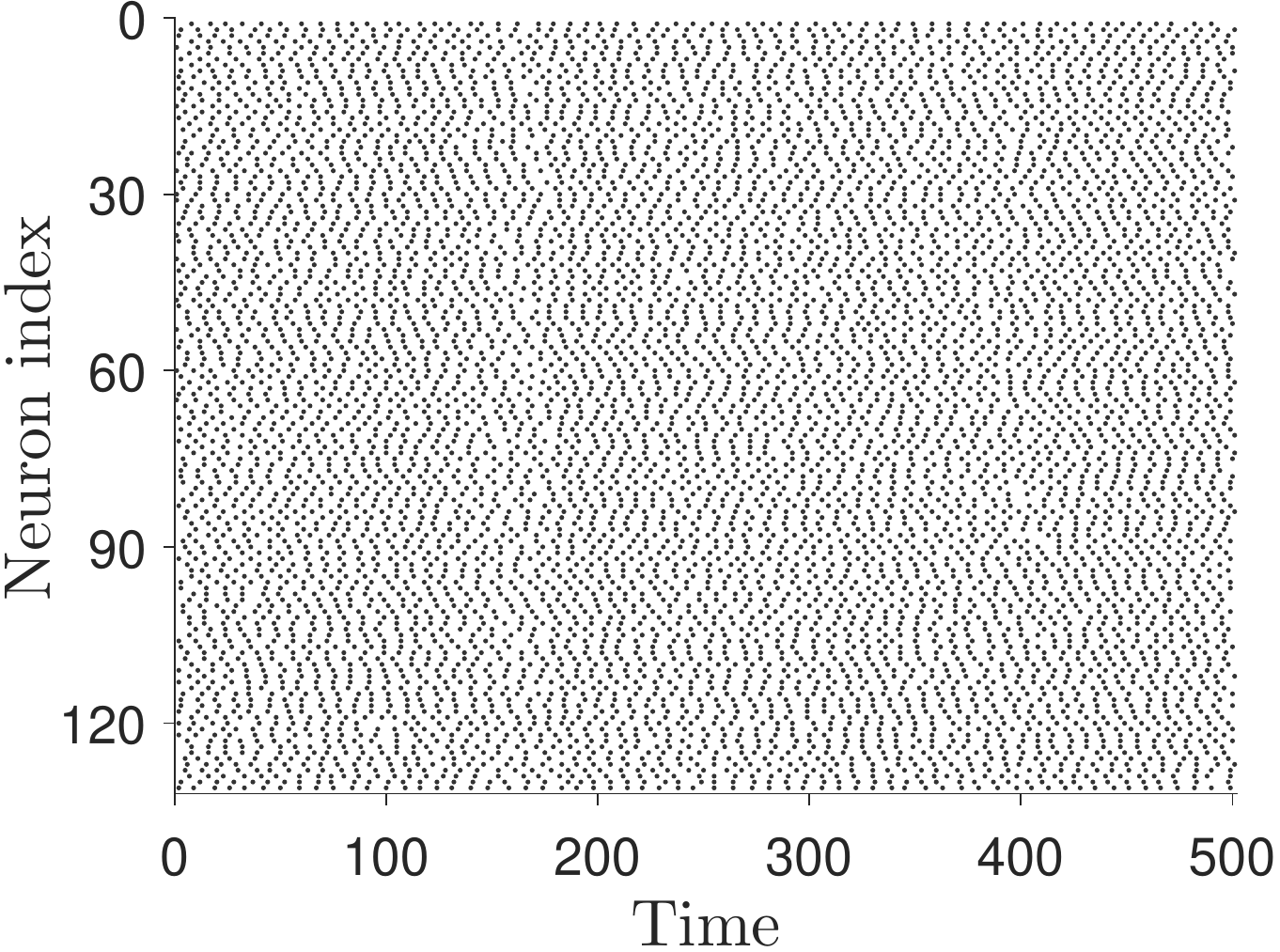}
   \end{center}
  \end{minipage}
 \end{center}

    \caption{\textbf{Collective dynamics for different coupling strengths and network structures}. In panel {\bf(a)} we show the ensemble-averaged order parameter, $R$, for the inter-spike intervals time-series of $N=131$ maps connected using the C. elegans frontal neural network (CE, with small-worldness coefficient $\sigma = 2.8$), Erd{\"o}s-R{\'e}nyi (ER, with $\sigma = 1.0$) and Watts-Strogatz (WS, with $\sigma=2.8$) ensembles. For the ER networks, panels {\bf (b)} ($\epsilon=0.23$) and {\bf (c)} ($\epsilon=0.26$) show raster plots indicating the firing pattern of the coupled neuron maps before and after the abrupt drop in panel {\bf (a)}'s $R$ values. Panel {\bf (d)} shows $R$ for the rewired ER and WS networks, such that all these ensembles have $\sigma = 2.1$ (for comparison, black dots show $R$ for the CE network). Panels {\bf (e)} ($\epsilon=0.23$)  and {\bf (f)} ($\epsilon=0.26$) show the corresponding raster plots.}
 \label{FIG:Collective_dynamics}
\end{figure}

The coupling strength $\epsilon^\ast \simeq 0.26$, which maximises the $TPR$, implies an average impulse per map of $\epsilon^\ast/\!\left\langle k \right\rangle \simeq 0.025$. This coupling is associated to a collective regime with a loosely coupled dynamic, as we show in Fig.~\ref{FIG:Collective_dynamics}, since it corresponds to an average $2.5\%$ synaptic increment of the membrane potential range (i.e., the difference between maximum and minimum membrane potential values) due to the action from neighbouring maps. The fact that our inference method recovers approximately $50\%$ of all existing connections at $\epsilon \sim 0.26$ ($TPR = 0.5$), implies that the $FNR = 0.5$, and from Eq.~\eqref{eq_TNR}, this also implies that $TNR \simeq 0.96$ and that $FPR \simeq 0.04$ for $\rho \simeq 0.08$. This means that for $\epsilon \sim 0.26$, the inference method is highly efficient in detecting true nonexistent connections ($TNR \to 1$) and falsely classifying these connections as existing ones ($FPR \to 0$). This efficiency is a consequence of sparse networks having more non-existing connections than existing connections ($N\,(N - 1)/2 \gg M$); as it happens in our ensembles. Hence, in sparse networks the challenge is to correctly identify the existing connections.

In spite of the similar results in Fig.~\ref{FIG:TPR_vs_Sg} panels, we can distinguish a slight advantage in the ER networks ensemble-averaged $TPR$ values [Fig.~\ref{FIG:TPR_vs_Sg}{\bf (a)}] over the WS $TPR$ values [Fig.~\ref{FIG:TPR_vs_Sg}{\bf (b)}], which also appear when $N = 277$. We understand that differences in the inference results have to appear because of the dependence on the underlying degree distribution (as well as in the small-worldness), as it is observed in the results for the CE and ER networks shown in Fig.~\ref{FIG:TPR_vs_Ep_CE_WS_ER}. In Fig.~\ref{FIG:TPR_vs_Sg}, the degree distributions correspond to those of the ER and WS networks respectively, which are kept invariant as the small-worldness of the underlying network is changed. However, the similarity in the results from Fig.~\ref{FIG:TPR_vs_Sg}{\bf (a)} and {\bf (b)} (for similar small-worldness values) can be explained due to the finite size systems, which make the ER and WS degree distributions similar (a similar behaviour is also observed for $N = 277$ -- not shown here). We also note that the modified ER networks $TPR$ results narrowly outperform WS inference results, where success rates reach values higher than $50\%$ for coupling strengths close to $\epsilon^\ast \simeq 0.26$. These $TPR$ values are significantly higher than making a blind random inference of connections (dashed horizontal lines), which successfully recover only $\simeq 8\%$ of the existing connections.

All $TPR$ curves share an abrupt increase in the success rate around a critical coupling strength of $\epsilon^* \approx 0.26$. Figures~\ref{FIG:TPR_vs_Ep_CE_WS_ER} and \ref{FIG:TPR_vs_Sg} show this abrupt jump in the inference success for all networks analysed -- though the exact value of $\epsilon^*$ may vary slightly for each topology realisation. This sudden increase in the success rates points to a drastic change in the systems' collective dynamics as the coupling strength is increased beyond $\epsilon^*$. In order to analyse the maps' collective dynamics as a function of $\epsilon$ and the underlying topology, we compute the order parameter, $R$, defined as the time-average of the squared difference between two inter-spike intervals (ISIs) time-series (i.e., the series of time differences between two consecutive spikes) summed over all pairs of ISIs. Specifically, $R = \sum_{j<i}{R_{ij}}$, with $R_{ij}= \langle(T_i -T_j)^2 \rangle_{t}$, where $T_i$ is the $i$-th neuron ISI time-series and $\langle \cdot \rangle_t$ is the time-average. This means that the $R$ value is high [low] when the time series are different [similar].

In Fig.~\ref{FIG:Collective_dynamics}{\bf (a)} we show how the order parameter $R$ changes with the coupling strength, $\epsilon$, for different systems with $N = 131$ neurons. Namely, the results for the C. elegans frontal neural network structure are shown by the filled (black online) circles, whereas for the ER and WS networks, the ensemble-averaged and initial-condition averaged $R$ values are shown by unfilled (blue online) diamonds and unfilled (red online) circles, respectively. The ensemble averages are found from $20$ realisations and from $10$ different initial conditions for each topology realisation. For the CE network each $R$ value represents an average over $10$ initial conditions. We can observe that close to $\epsilon\approx 0.25$, $R$ decreases abruptly for all network structures falling $2$ orders in magnitude. This drop corresponds to a switch from a collective bursting regime to a lightly disordered spiking regime which has higher collective coherence. This spiking regime appears as disordered when compared to the apparently synchronous bursting dynamics, but it is in fact partially coherent -- a particularly suitable condition to perform a successful network inference \cite{rubido2014exact}. For example, Fig.~\ref{FIG:Collective_dynamics} panels {\bf (b)} and {\bf (c)} show the raster plots for ER networks with $N = 131$ maps coupled with $\epsilon=0.23$ and $\epsilon=0.26$, respectively. Similarly, Fig.~\ref{FIG:Collective_dynamics}{\bf (d)} shows the same behaviour for the averaged $R$ parameter in networks with ER and WS degree distributions but with different small-worldness levels (as previously described). This means that the collective dynamics' abrupt change also happens for the modified networks, namely, the networks modified by our rewiring process to increase or decrease their overall small-world coefficient. Panels \textbf{(e)} and \textbf{(f)} show the resultant raster plots for $\epsilon=0.23$ and $0.26$, respectively, for a realisation of an ER network with $N=131$ maps and $\sigma=2.1$.

Furthermore, we can see from Figs.~\ref{FIG:TPR_vs_Ep_CE_WS_ER} and \ref{FIG:TPR_vs_Sg} that the sensitivity falls rather smoothly for all networks as we increase $\epsilon$ beyond the critical value $\epsilon^*$. The reason behind this smooth change is that, as $\epsilon$ increases beyond $\epsilon^*$, the neurons gradually begin to fire in a more ordered spiking, namely, achieving synchronisation. Thus, partial coherence between the time-series vanishes and inference becomes impossible. The smooth decrease in sensitivity can be observed by the rate in which the order parameter decreases for $\epsilon$ larger than $\epsilon^*$, as in Figs.~\ref{FIG:Collective_dynamics} panels \textbf{(a)} and \textbf{(d)}. In general, the neural systems we analyse stay in a partially coherent spiking regime for an interval of coupling strength values (approximately between $\epsilon \approx 0.25$ and $0.30$), where network inference $TPR$ values remain above the random line.

\begin{figure}[htbp]

 \begin{center}
  \begin{minipage}{0.49\textwidth}
   \begin{center}
    {\bf (a)}\\ \includegraphics [scale= 0.58] {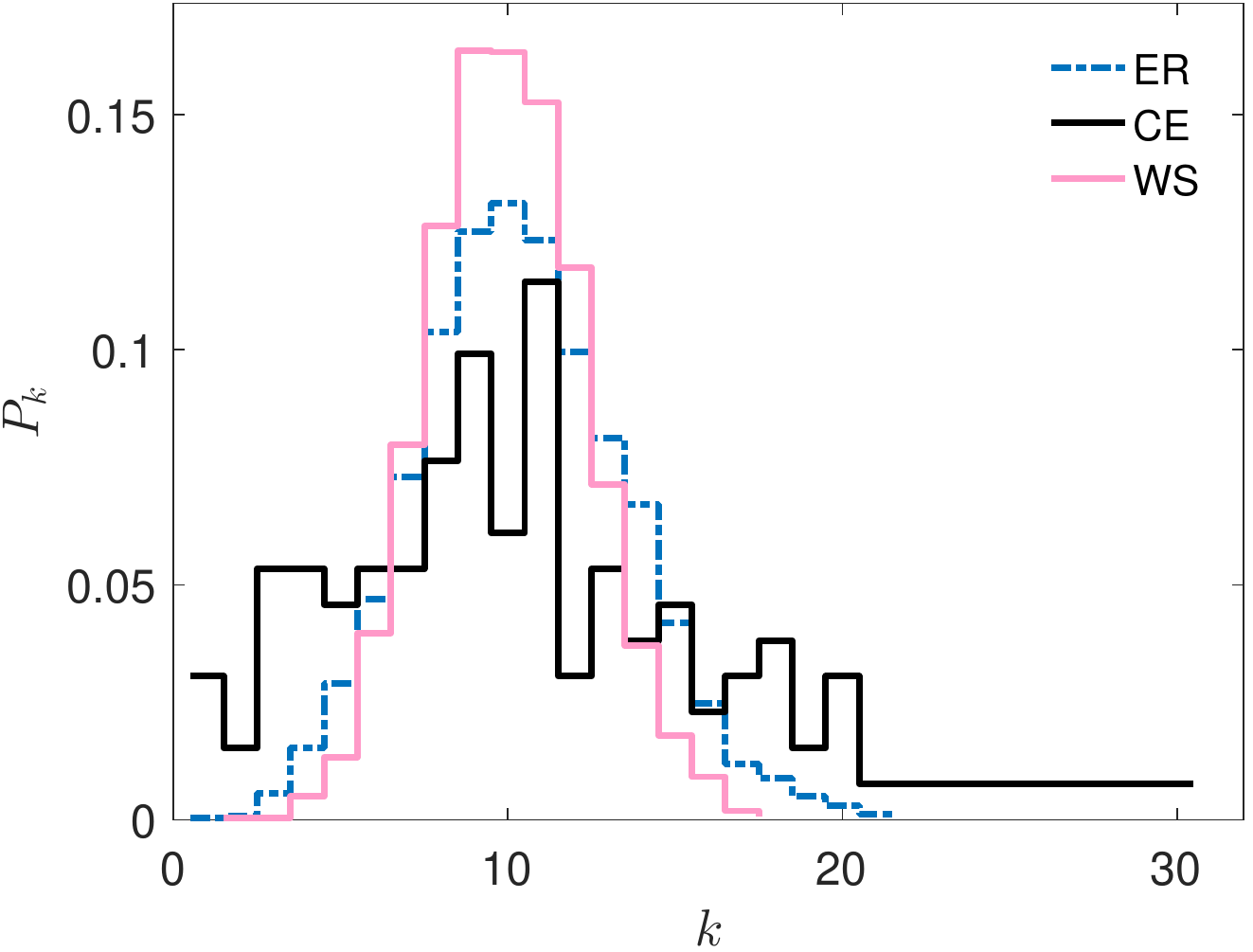}
   \end{center}
  \end{minipage} \hspace{0.2pc}
  \begin{minipage}{0.49\textwidth}
   \begin{center}
    {\bf (b)}\\ \includegraphics [scale= 0.58] {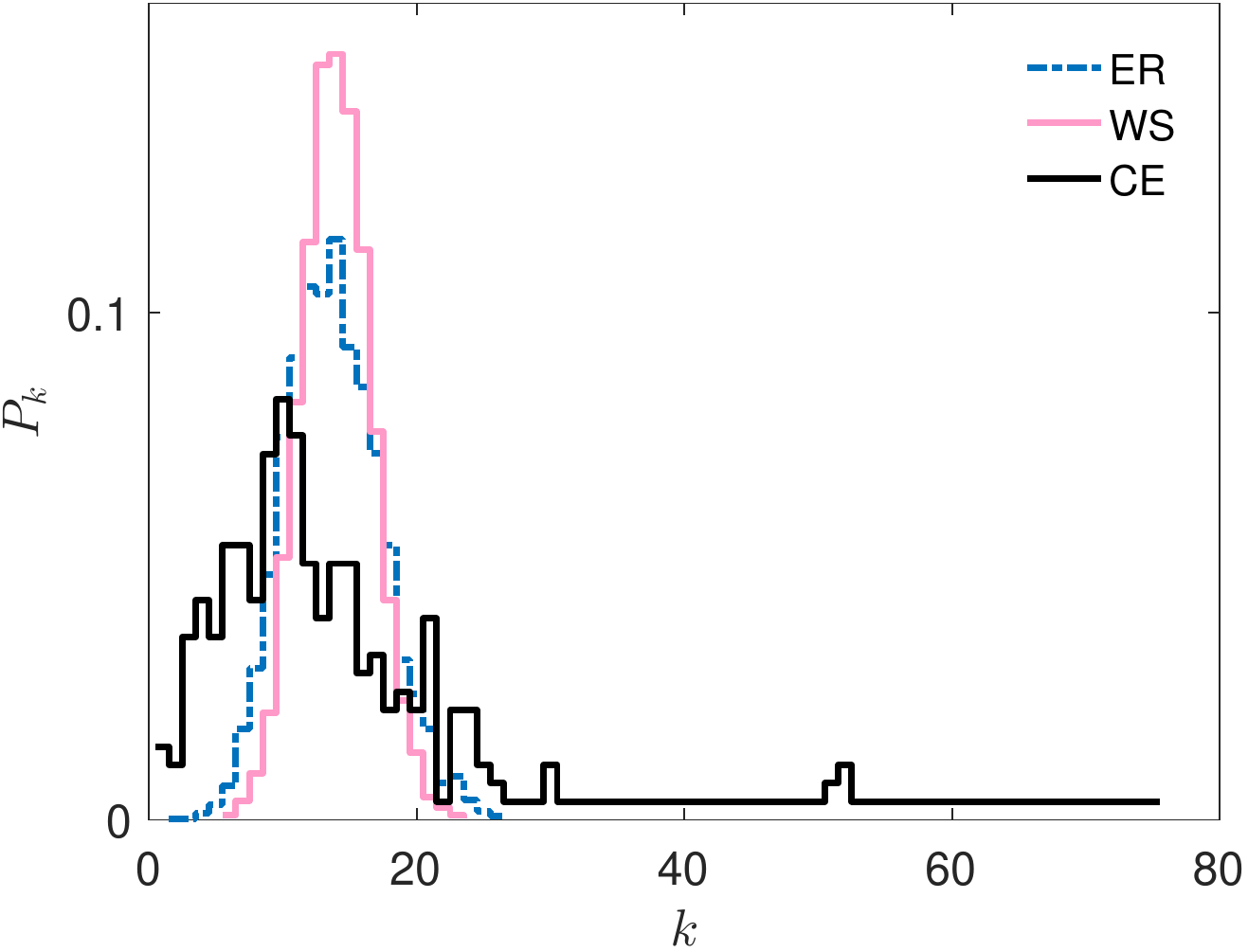}
   \end{center}
  \end{minipage}
 \end{center}
    \caption{\textbf{Average degree distributions of our neural network structures}. Panel {\bf (a)} [{\bf (b)}] shows the $N=131$ [$N=277$] nodes degree-distributions for Erd{\"o}s-R{\'e}nyi (ER, dashed -- blue online) and Watts-Strogatz (WS, continuous -- pink online) ensembles, averaged over $20$ network realisations. Also, the C. elegans (CE) frontal [global] neural network structure is shown with continuous black lines.}
 \label{FIG:Degree_dist}
\end{figure}

So far we have shown that increasing small-worldness favours network inference, obtaining success rates that appear robust to changes in the degree-distribution type (e.g., ER, WS, and CE), initial conditions, and similar for a broad coupling strength region. However, we can see from Fig.~\ref{FIG:TPR_vs_Ep_CE_WS_ER} that as the $N$ increases from $131$ (panel {\bf (a)}) to $277$ (panel {\bf (b)}), the $TPR$ drops significantly for the C. elegans networks. The reason for this drop comes from the broadness in the global CE neural-network's degree-distribution. As we can see from Fig.~\ref{FIG:Degree_dist}, when $N = 131$ (panel {\bf (a)}), all degree distributions are somewhat similar and narrow, but when $N = 277$ (panel {\bf (b)}), the CE topology shows the presence of hubs and a long tailed distribution. This is why on Fig.~\ref{FIG:TPR_vs_Ep_CE_WS_ER}{\bf (b)}, the $TPR$ results for the $N = 277$ WS network ensemble are extremely similar to those $TPR$ values when $N = 131$ in Fig.~\ref{FIG:TPR_vs_Ep_CE_WS_ER}{\bf (a)}. Similarly, we can see the same resemblance in the $TPR$ results for ER networks, which also hold a narrow degree distribution, as shown by the dashed curves in Fig.~\ref{FIG:Degree_dist}. On the contrary, the significant differences emerging from the CE degree distributions for $N = 131$ and $N = 277$ impact directly into the inference success rates. This leads us to believe that heterogeneity in the node degrees hinders network inference.

We find that the former results are also robust to changes in the time-series length. Moreover, our conclusions regarding small-worldness favouring inference still hold if one chooses a different time-series representation for the neural dynamics of each map. Namely, our findings hold for inter-spike intervals (ISI) as well as membrane potentials (see \href{sec_SuppMat}{Supplementary Information} for further details). Remarkably, we see that when using ISIs to measure the neural activity, inference success rates are significantly lower than when using membrane potentials -- regardless of the particular topology or collective dynamics. However, these $TPR$ values are more robust to changes in the topology realisation (i.e., different structures constructed with the same canonical models and parameters). This leads us to conclude that, using ISIs instead of membrane potentials, allows us to achieve worst inference success rates but with a more reliable outcome.

\section*{Discussion}
In this work we have shown that the network inference methods based on the use of cross-correlation (CC) to measure similarities between components are more effective when inferring small-world structures than other types of networks. This conclusion has broad implications, since cross-correlation is widely used to reveal the underlying connectivity of neural systems, such as the brain, and to gain information about long-range interaction in climate networks.

We have shown that, for networks with similar degree distributions, the small-worldness level is the main topological factor affecting the inference success rates. The results shown in Figs.~\ref{FIG:TPR_vs_Ep_CE_WS_ER} and \ref{FIG:TPR_vs_Sg} account for the reliability and robustness of this conclusion. Also, in Fig.~\ref{FIG:TPR_vs_Ep_CE_WS_ER} we observe that our inferring method is consistently and robustly more successful when inferring Watts-Strogatz networks than the C. elegans (CE) neural structure, despite both having the same small-worldness level. This points to the effect that degree-distribution range has on the inference success, namely, the broader the degree distribution, the less effective the inference is.

Our results show that the appearance of highly connected nodes (hubs), such as in the CE global network, is another important factor, hindering successful inference. In other words, we find that success rates are generally lower when inferring networks which have higher degree heterogeneity. This finding is relevant because real small-world networks, such as the C. elegans neural structure, often combine small-world properties with other complex features -- such as the presence of hubs, hierarchies or rich clubs --, resulting in a higher degree heterogeneity with respect to the canonical Watts-Strogatz network model. In particular, hubs have been related to the phenomenon of hub synchronisation in brain networks \cite{vlasov2017hub} and scale-free networks\cite{pereira2010hub}. Hub synchronisation is particularly detrimental for network inference, since it leads to strong correlations between non-connected nodes and weak correlations between the hubs and their neighbouring nodes.

Regardless of the underlying structure, the coupling strength range that allows for a successful network inference is located after a critical value, in which the system transitions from a collective dynamics with an apparently synchronous bursting regime to collective asynchronous spiking. This transition has been reported to take place in C. elegans neural networks\cite{protachevicz2019bistable} and corresponds to a partially-coherent state, which has been shown to be necessary for having successful network inference\cite{rubido2014exact}. Figure~\ref{FIG:Collective_dynamics} shows an abrupt change in the systems' order parameter around the critical coupling strength, revealing a coherence-loss after the transition. Although we can only perform successful network inferences when the systems are in partial coherence, it is reasonable to assume that real neural systems are in such states, since they consistently transition between synchronous and asynchronous states in order to perform different tasks and cognitive functions \cite{bassett2017network,medaglia2015cognitive,sporns2014contributions,bullmore2009complex,sporns2005human,deco2015rethinking,protachevicz2019bistable,hizanidis2016chimera}.


\section*{Methods}
 \label{sec_methods}
 
    \subsection*{Synthetic neural network model}
Our synthetic neural model is the Izhikevich map \cite{izhikevich2003simple,izhikevich2004model,ibarz2011map}, which belongs to the bi-dimensional quadratic integrate-and-fire family. This map consists of a fast variable, $v$, representing the membrane potential, and a slow variable, $u$, modelling changes in the conductance of the ionic channels. One of the main advantages of using Izhikevich maps is that it combines numerical efficiency (inherent to map-based models) with biological plausibility\cite{izhikevich2004model}. The isolated map equations of motion are given by
\begin{equation}
 \left\{ \begin{array}{l}
     v_{n+1}=0.04v_{n}^{2}+6v_{n}+140+I-u_{n} \\
     u_{n+1}=0.02(0.25v_{n}-u_{n})+u_{n}
 \end{array}, \right. \;\;\text{if}\;\; v_{n} < 30\,mV,\;\;\;\text{and}\;
 \left\{ \begin{array}{l}
    v_{n+1}=c  \\
    u_{n+1}=u_{n}+d
 \end{array}, \right.\;\;\text{if}\;\; v_{n} \geq 30\,mV.
    \label{eq:Izhikevich_isolated}
\end{equation}

\noindent When different values of $I$, $c$, and $d$, are fixed, the Izhikevich map can show extensive dynamical regimes, which have been observed in real neurons. We set the parameter values such that the regime exhibits bursting dynamics. Namely, $d = 0$, $c = -58$, and $I=2$. However, when Izhikevich maps interact, the resulting single-neuron dynamics can differ significantly from the bursting regime.

The interactions are set to be pulsed via the fast variable, $v$, and controlled by a global coupling-strength order parameter, $\epsilon$. This pulse-coupling type is able to represent many real neural interactions. With this model, every time a neuron spikes it sends a signal to the adjacent neurons (i.e., to all neurons that are connected to it), instantly advancing their membrane potentials by a constant value. Specifically, the dynamics for the $n$-th neuron is given by
\begin{eqnarray}\nonumber
 \left\{ \begin{array}{l}
      v_{i,n+1}=0.04v_{i,n}^{2}+6v_{i,n}+140+I-u_{i,n}+\frac{\epsilon}{k_{i}}\sum_{j\neq i}A_{ij}\delta(v_{j,n}-30) \\
      u_{i,n+1}=a(bv_{i,n}-u_{i,n})+u_{i,n}
 \end{array} \right., \;\;\text{if}\;\; v_{i,n} < 30\,mV, \;\;\;\;\text{and}\\
 \left\{ \begin{array}{l}
      v_{n+1}=c  \\
      u_{n+1}=u_{n}+d
 \end{array} \right., \;\;\text{if}\;\; v_{i,n} \geq 30\,mV,
    \label{eq:Coupled_neuron_dynamics}
 \end{eqnarray}
where $k_{i}$ is the $i$-th node degree (i.e., its number of neighbours), $A_{ij}$ is the $ij$-th entry of the network's adjacency matrix, $\epsilon$ is the coupling strength, and $\delta(x)$ is the Kronecker's delta-function. We iterate Eq.~\eqref{eq:Coupled_neuron_dynamics} $7\times10^4$ steps from $10$ random initial conditions for each topology and coupling strength, removing a transient of $2\times10^4$ steps. In particular, the coupling term, $\frac{\epsilon}{k_{i}}\sum_{j\neq i}A_{ij}\delta(v_{j,n}-30)$, acts as follows. If a connection between neurons $i$ and $j$ exists, then $A_{i,j} = 1$ and neuron $i$ receives an input of value $\epsilon/k_{i}$ every time neuron $j$ reaches the threshold $v_{j,n}=30$. Otherwise, the neuron remains unchanged.

    \subsection*{C. elegans neural structure}
We use data from Dynamic Connectome Lab \cite{varier2011neural,ren2010stdp} to construct the C. elegans' frontal ($N=131$ nodes) an global ($N=277$ nodes) neural networks. These networks are represented by weighted and directed graphs, that we simplify by considering the unweighted non-directed versions. The reason behind this choice is to bring up-front the role of structure alone into our systems' collective behavior. Under these symmetric considerations, we find that the network's mean degree is $\overline{k}_f = 10.5$ ($\overline{k}_g = 13.8$) for the frontal (global) connectome, with a sparse edge density of $\rho_f =0.08$ ($\rho_g = 0.05$). The average shortest-path length for our C. elegans frontal (global) neural network is $\overline{l}_f= 2.5$ ($\overline{l}_g=2.6$) and its clustering coefficient is $C_f=0.25$ ($Cg=0.28$). Both neural structures have short average path lengths (similar to Erdös-Rényi networks with equal edge density) and high clustering coefficients (similar to Watts-Strogatz networks with equal average degree), hence, they show small-world properties. The small-worldness coefficient of our C. elegans frontal (global) neural network is $\sigma_f=2.8$ ($\sigma_g=5.1$), which falls within the expected small-world range ($\sigma>2$).

\subsection*{Network ensembles}
In order to study the role that small-worldness and degree heterogeneity have in the network inference results, we build two ensembles of $20$ Erd{\"o}s-R{\'e}nyi (ER) adjacency matrices with $N=131$ and $N=277$ nodes respectively, and two ensembles of $20$ Watts-Strogatz (WS) adjacency matrices. We choose the number of nodes in our network ensembles to match the sizes of the C. elegans (CE) frontal ($N=131$ nodes) and global ($N=277$) neural structures. We also tune the algorithms to build these networks such that they have the same edge densities as the CE frontal and global neural networks, namely, $\rho=0.08$ and $\rho=0.05$, respectively. In addition, the WS ensemble is also tuned such that the algorithm parameters produce networks with similar small-worldness levels to that of the CE networks. In what follows, $\langle k \rangle$ denotes average among network ensembles, while $\overline{k}$ expresses the average among nodes of a single network. 

Our ER ensemble is built with a probability to linking nodes in each network of $p=0.08$ ($0.05$) for $N=131$ ($N=277$) -- values which are above the percolation transition. These probabilities yield mean degrees of $\langle \overline{k} \rangle=10.5$ ($\langle \overline{k}\rangle=13.8$) for the $N=131$ ($N=277$) networks. In both cases, the variability within the ensemble of these mean degrees is $\sigma_{\overline{k}}=0.3$. We can corroborate that the nematode's neural networks also have mean degrees falling within one standard deviation of the ER ensemble-averaged mean degrees. The clustering coefficient in the ER model is usually low ($C=p$ in the thermodynamic limit), being $\langle C \rangle =0.08$ ($\langle C \rangle=0.05$) for our $N=131$ ($N=277$) network ensembles. The ER networks also hold a small average shortest-path length. In our ensembles, the shortest-path lengths are $\langle \overline{l} \rangle=2.3$ ($\langle \overline{l} \rangle=2.4$) for the $N=131$ ($N=277$) networks. Correspondingly, the averaged small-worldness levels of our ER ensembles are $\langle \sigma \rangle=1.0$ in both cases -- as expected --, indicating the absence of small-world effect.

The Watts-Strogatz (WS) algorithm takes an initial ring configuration in which all nodes are linked to $K/2$ neighbours to each side, and then rewires all edges according to some probability $p$  \cite{watts1998collective}. Using this model, we construct a network ensemble with link density and small-worldness levels similar to the CE neural networks. In particular, we choose a mean degree and rewiring probability that yields similar average path lengths and clustering coefficients as the CE neural structures. Specifically, we fix the mean degree at an integer value, namely, $ \langle k \rangle = 10$ ($14$) for $N=131$ ($277$) nodes. Then, for each rewiring probability $p$, we generate $20$ adjacency matrices and calculate the mean average path length and clustering coefficients. Thus, for each rewiring probability $p$ we have a point in the $[C, \langle l \rangle]$ space. This allows us to choose the rewiring probability, $p^\star$, which holds the  closest point in the $[ C , \langle l \rangle]$ space to the CE networks values. For such $p^\star$, our ensembles have shortest-path lengths of $\langle \overline{l} \rangle = 2.5$ ($\langle \overline{l} \rangle = 2.6$), and clustering coefficients of $C=0.4$ ($C=0.28$) when $N=131$ ($N=277$). Correspondingly, our networks' small-worldness levels are $\langle \sigma \rangle = 2.8$ ($\langle \sigma \rangle=5.3$) when $N=131$ ($N=277$). These values are similar to the CE small-worldness levels and indicate the presence of the small-world effect.


\bibliography{references}

\section*{Acknowledgements}
R.A.G and N.R. acknowledge Comisi{\'o}n Sectorial de Investigaci{\'o}n Cient{\'i}fica (CSIC) research grant 97/2016 (ini\_2015\_nomina\_m2). All authors acknowledge CSIC group grant "CSIC2018 - FID13 - grupo ID 722".     

\section*{Author contributions statement}
N.R. conceived the experiments. R.A.G. conducted the experiments All authors analysed the results and contributed to the writing of the manuscript.

\section*{Additional information}
\textbf{Competing financial interests: } the authors declare no competing financial interests.

\newpage

\section*{Supplementary Information}\label{sec_SuppMat}

\subsection*{Network inference using interspike intervals}

\begin{figure}[htbp]
 \begin{center}
  \begin{minipage}{0.49\textwidth}
   \begin{center}
    {\bf (a)}\\ \includegraphics[scale= 0.62]{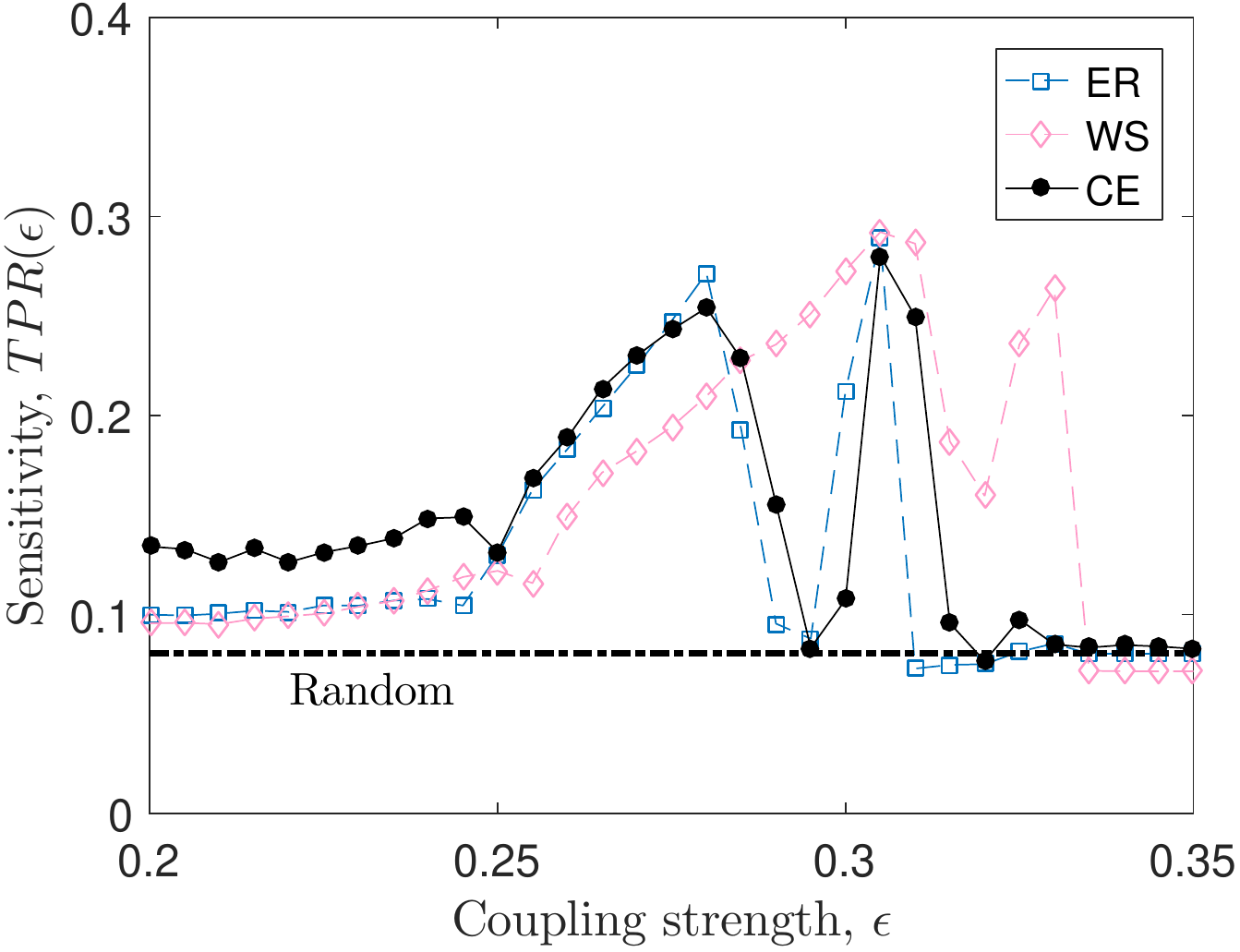}
   \end{center}
  \end{minipage} \hspace{0.2pc}
  \begin{minipage}{0.49\textwidth}
   \begin{center}
    {\bf (b)}\\ \includegraphics[scale= 0.62]{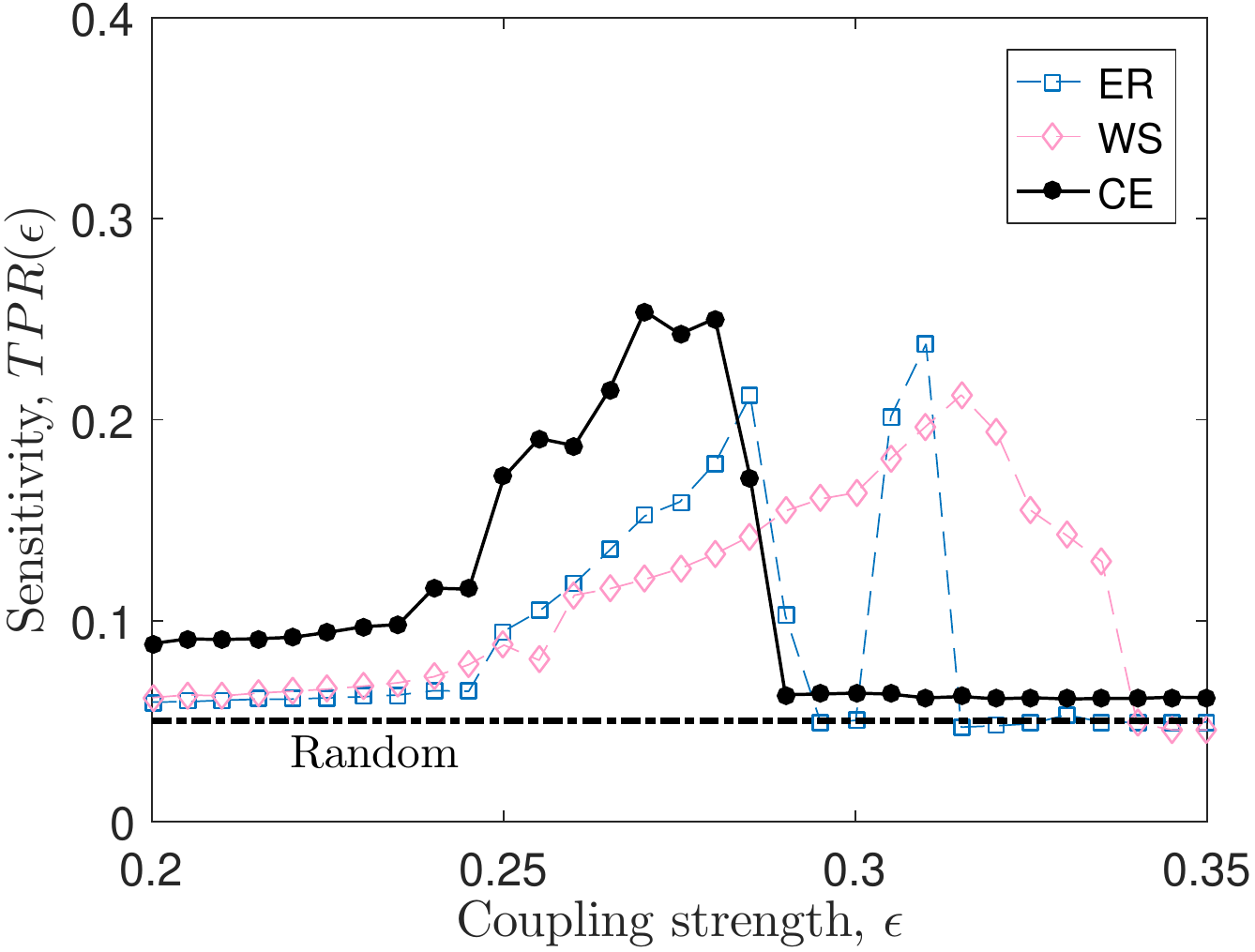}
   \end{center}
  \end{minipage}
 \end{center}
    \caption{{\bf Network inference success rates for different networks, coupling strengths, and sizes, using the interspike intervals time-series.} Panel {\bf (a)} [Panel {\bf (b)}] shows the true positive rate, $TPR$, as a function of the coupling strength, $\epsilon$, for $N = 131$ [$N = 277$] pulse-coupled Izhikevich maps connected in Erd{\"o}s-R{\'e}nyi (ER), Watts-Strogatz (WS), or C. elegans (CE) frontal [global] neural networks; with map parameters set such that the isolated dynamics is bursting (see \href{sec_methods}{Methods}). The $TPR$ values for the ER and WS come from averaging the results over $10$ initial conditions and $20$ network realisations with similar topological properties to that of the CE (i.e., number of nodes, average degree, and density of connections). For the CE, the results are averaged only on the initial conditions. The $TPR$ in each case is found by the same process as Fig.~\ref{FIG:TPR_vs_Ep_CE_WS_ER}, but using the interspike intevals time-series (considering $2000$ spikes) instead of the membrane potentials. The horizontal dashed line in both panels is the random inference $TPR$, namely, the null hypothesis.} 
 \label{FIG:TPRISI_vs_Ep_CE_WS_ER}
\end{figure}

The results for the network inference processes shown in Figs.~\ref{FIG:TPR_vs_Ep_CE_WS_ER} and \ref{FIG:TPR_vs_Sg} use the membrane potentials time-series to represent neural behaviour. Alternatively, many studies represent neural behaviour by a reduced time-series, the \textit{interspike intervals} \cite{sauer1994reconstruction,reinoso2016emergence}, which are the series of time intervals between two consecutive spikes, arguing that the most relevant information about neural behaviour is coded in this reduced representation of the dynamics. Hence, it is relevant to assess how our results are affected by different representations of neural behaviour. Fig.~\ref{FIG:TPRISI_vs_Ep_CE_WS_ER} shows the sensitivity as a function of the coupling strength for Erd{\"o}s-R{\'e}nyi, Watts-Strogatz and C. elegans networks under the same conditions as Fig.~\ref{FIG:TPR_vs_Ep_CE_WS_ER}, but in this case the cross-correlations are calculated between the interspike intervals series instead of the membrane potentials. Comparing Figs.~\ref{FIG:TPRISI_vs_Ep_CE_WS_ER} and \ref{FIG:TPR_vs_Ep_CE_WS_ER} we can observe that, in general, the sensitivity is lower when the network inference process is done using the interspike intervals than when we use the full information from the membrane potentials. Thus, the information reduction implied in the use of the interspike intervals to represent neural behaviour affects negatively the efficiency of our network inference processes.

\begin{figure}[htbp]
 \begin{center}
  \begin{minipage}{0.49\textwidth}
   \begin{center}
    {\bf (a)}\\ \includegraphics[scale= 0.62]{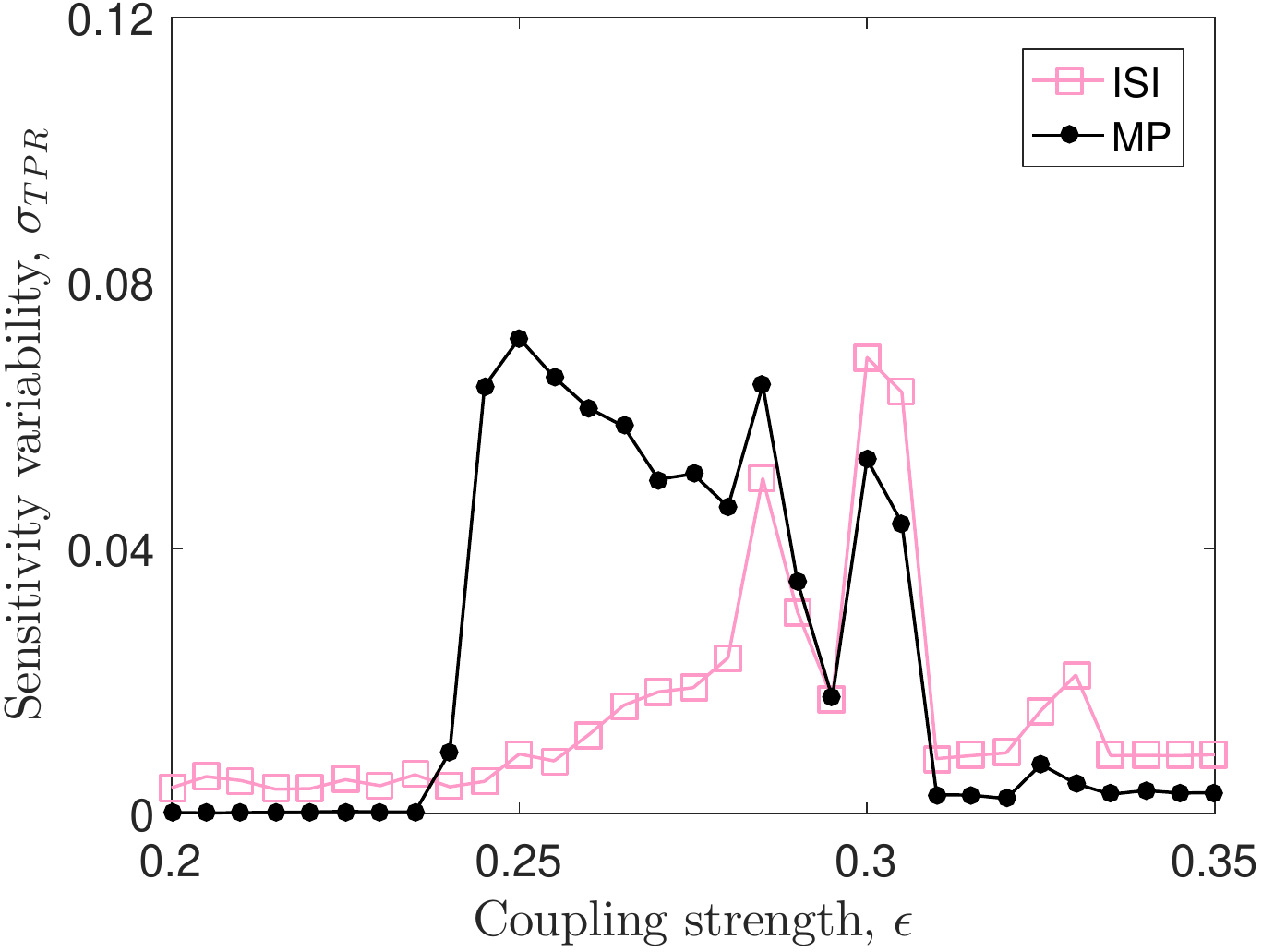} \\ 
    {\bf (c)} \\ \includegraphics[scale= 0.62]{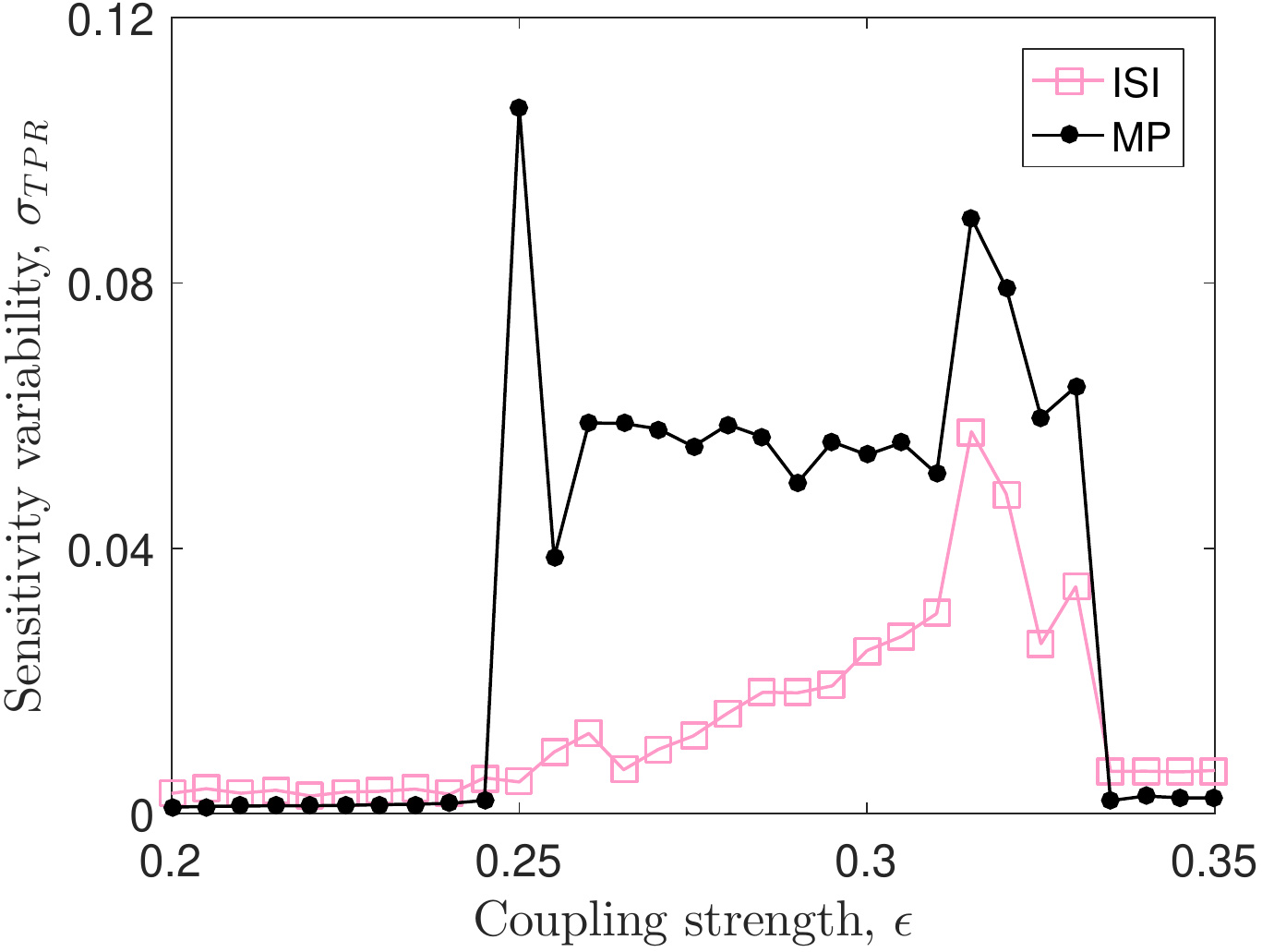} 
   \end{center}
  \end{minipage} \hspace{0.2pc}
  \begin{minipage}{0.49\textwidth}
   \begin{center}
    {\bf (b)}\\ \includegraphics[scale= 0.62]{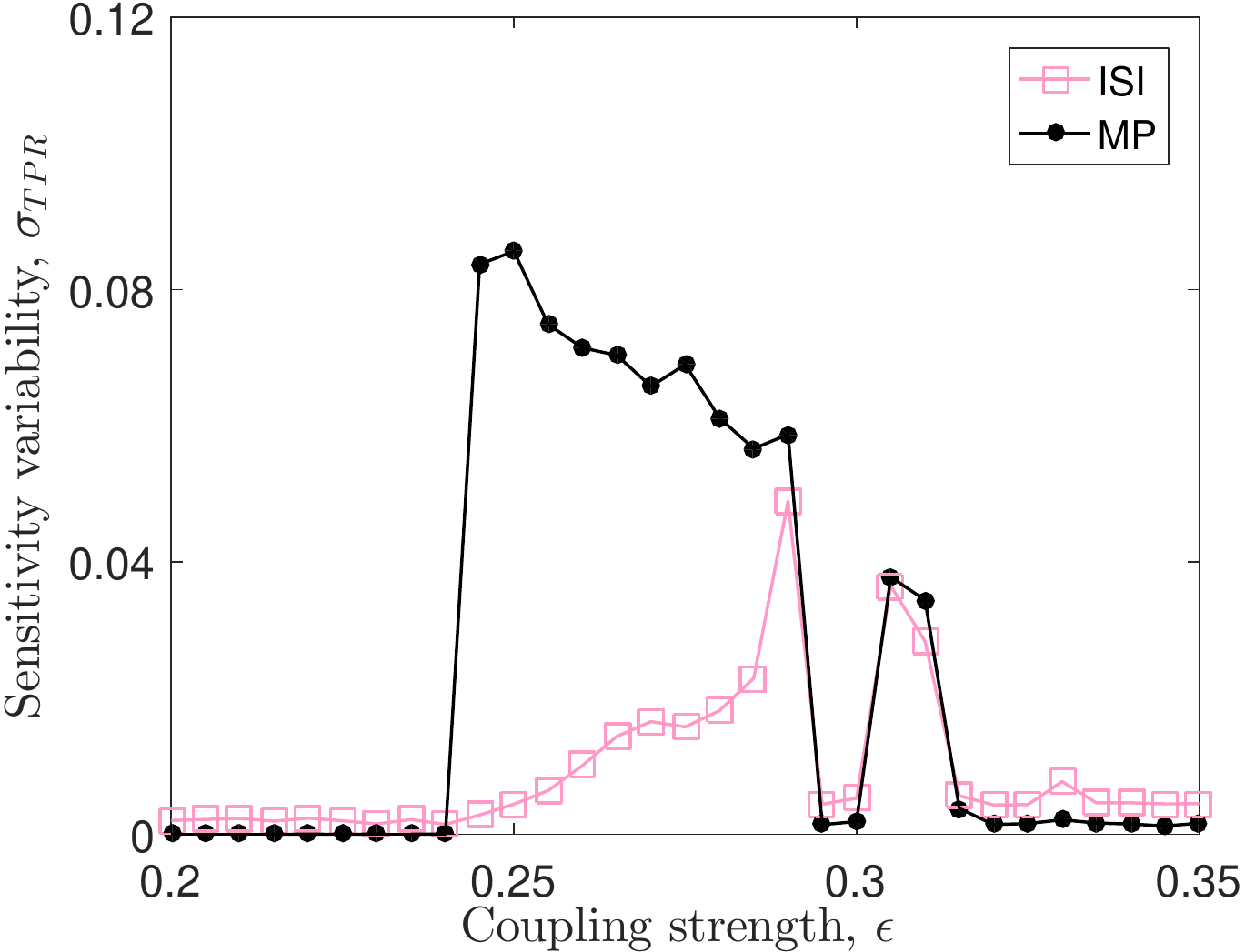} \\ 
    {\bf (d)} \\ \includegraphics[scale= 0.62]{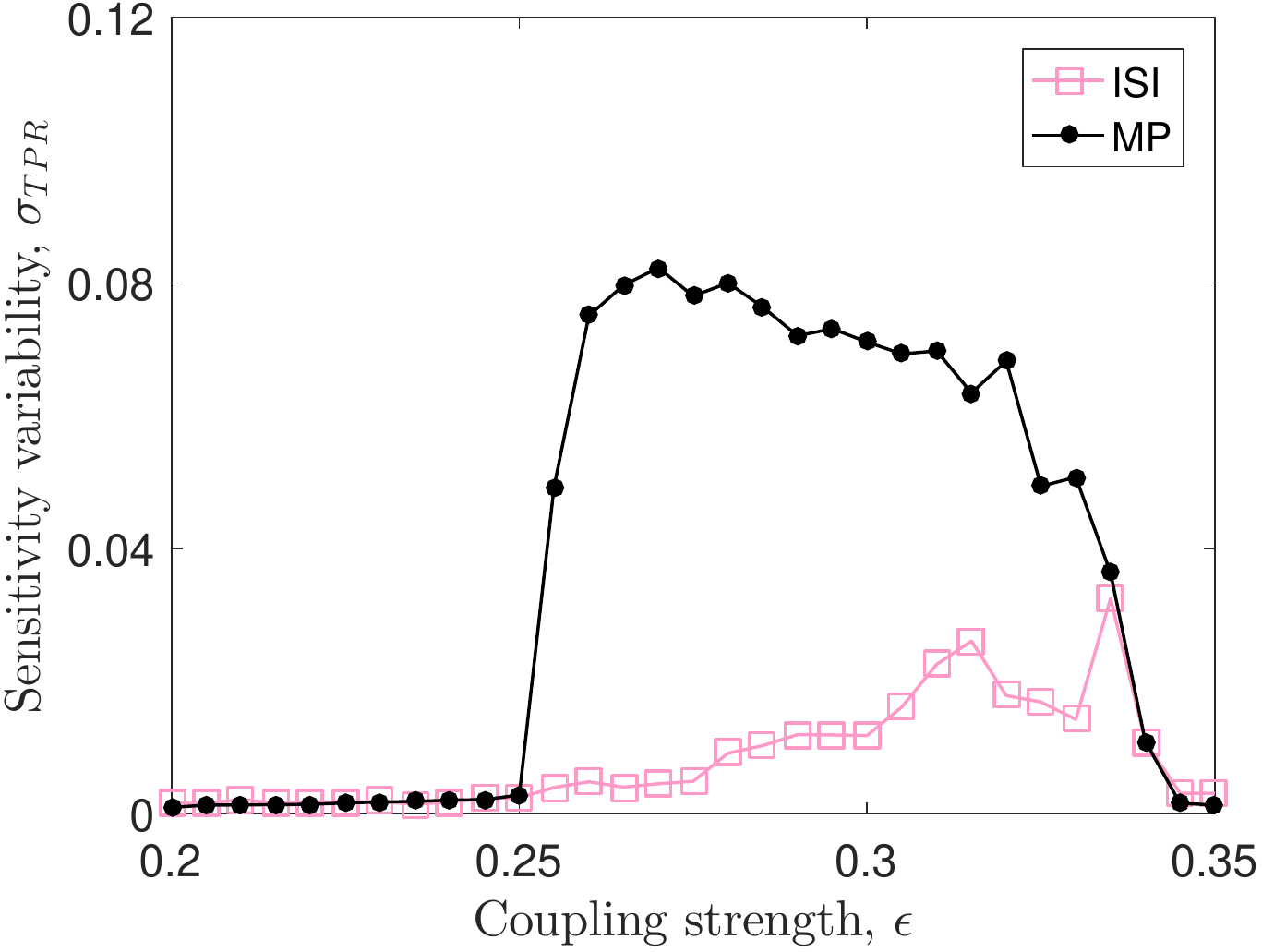} 
   \end{center}
  \end{minipage}
 \end{center}
    \caption{{\bf Standard deviation of the sensitivity as a function of the coupling strength in the Erd{\"o}s-R{\'e}nyi and Watts-Strogatz network ensembles.} In panels \textbf{(a)} and \textbf{(c)} [\textbf{(b)} and \textbf{(d)}] we show the standard deviation, $\sigma_{TPR}$, as a function of the coupling strength, $\epsilon$, in the $N=131$ [$N=277$] ER and WS network ensembles, respectively. All network ensembles consist of $20$ adjacency matrices, and all points result from an average over $10$ different initial conditions.} 
 \label{FIG:STD_TPR}
\end{figure}

In order to gain insight on the differences between the use of interspike intervals and membrane potentials to perform network inference process, another key factor is the robustness of the inference sensitivity among different realisations of the topology. We can assess this aspect by comparing the standard deviation of the sensitivity, $\sigma_{TPR}$, in our $N=131$ and $N=277$ ER and WS network ensembles, when we perform inference processes using membrane potentials and interspike intervals. Fig.~\ref{FIG:STD_TPR} shows $\sigma_{TPR}$ as a function of coupling strengths in the $N=131$ and $N=277$ ER and WS network ensembles. We observe that, in all cases, the sensitivity's standard deviation is considerably higher when using membrane potentials than when employing interspike intervals, for most coupling strengths. Combining these results with those shown in Fig.~\ref{FIG:TPRISI_vs_Ep_CE_WS_ER}, we conclude that while using the interspike intervals (instead of the full membrane potentials information) hinders our network inference process, it provides a higher robustness. Hence, there is a trade-off between efficiency and robustness when choosing different neural behaviour representations.        

\subsection*{Different time-series length}

\begin{figure}[htbp]
 \begin{center}
  \begin{minipage}{0.49\textwidth}
   \begin{center}
    {\bf (a)}\\ \includegraphics[scale= 0.62]{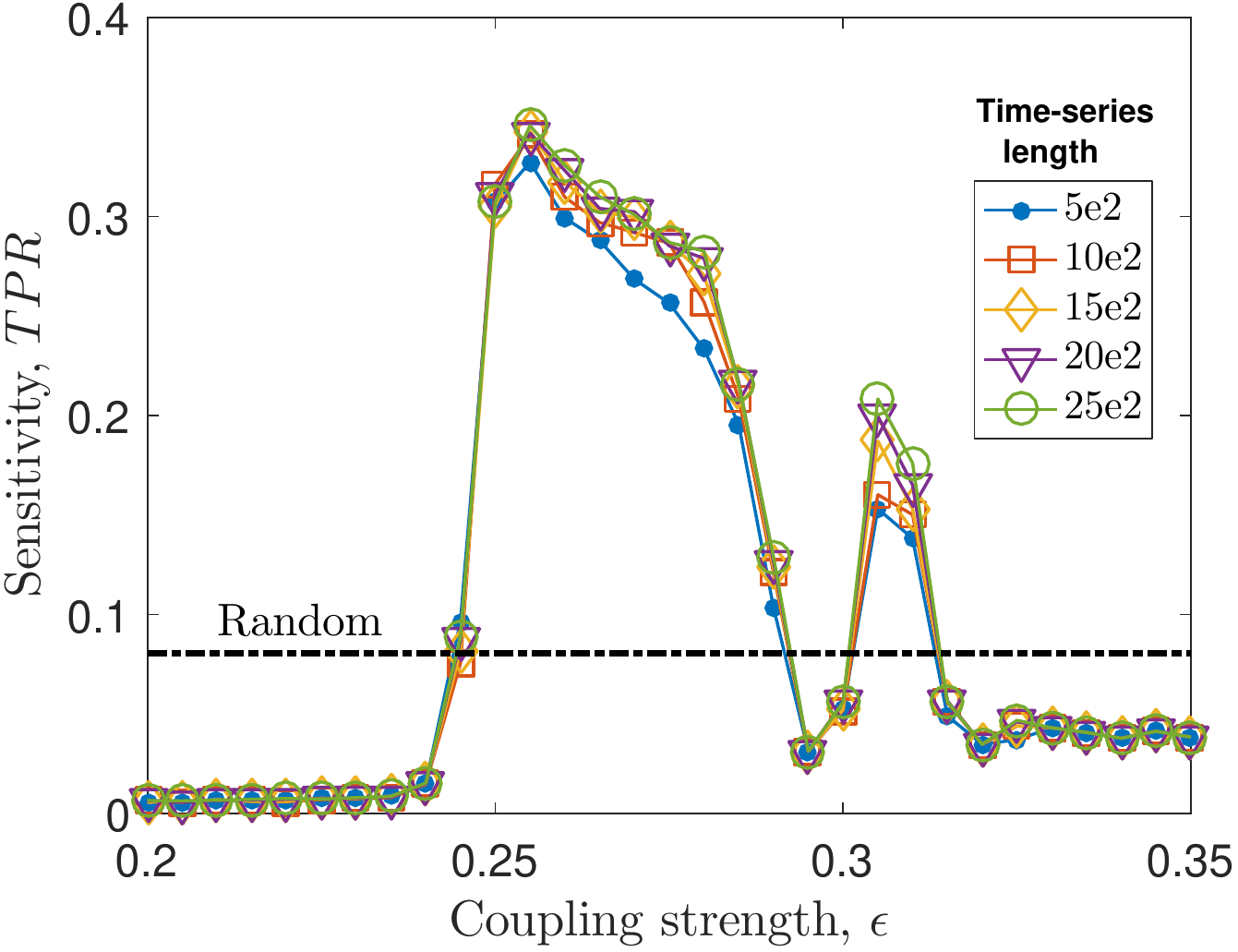}
   \end{center}
  \end{minipage} \hspace{0.2pc}
  \begin{minipage}{0.49\textwidth}
   \begin{center}
    {\bf (b)}\\ \includegraphics[scale= 0.62]{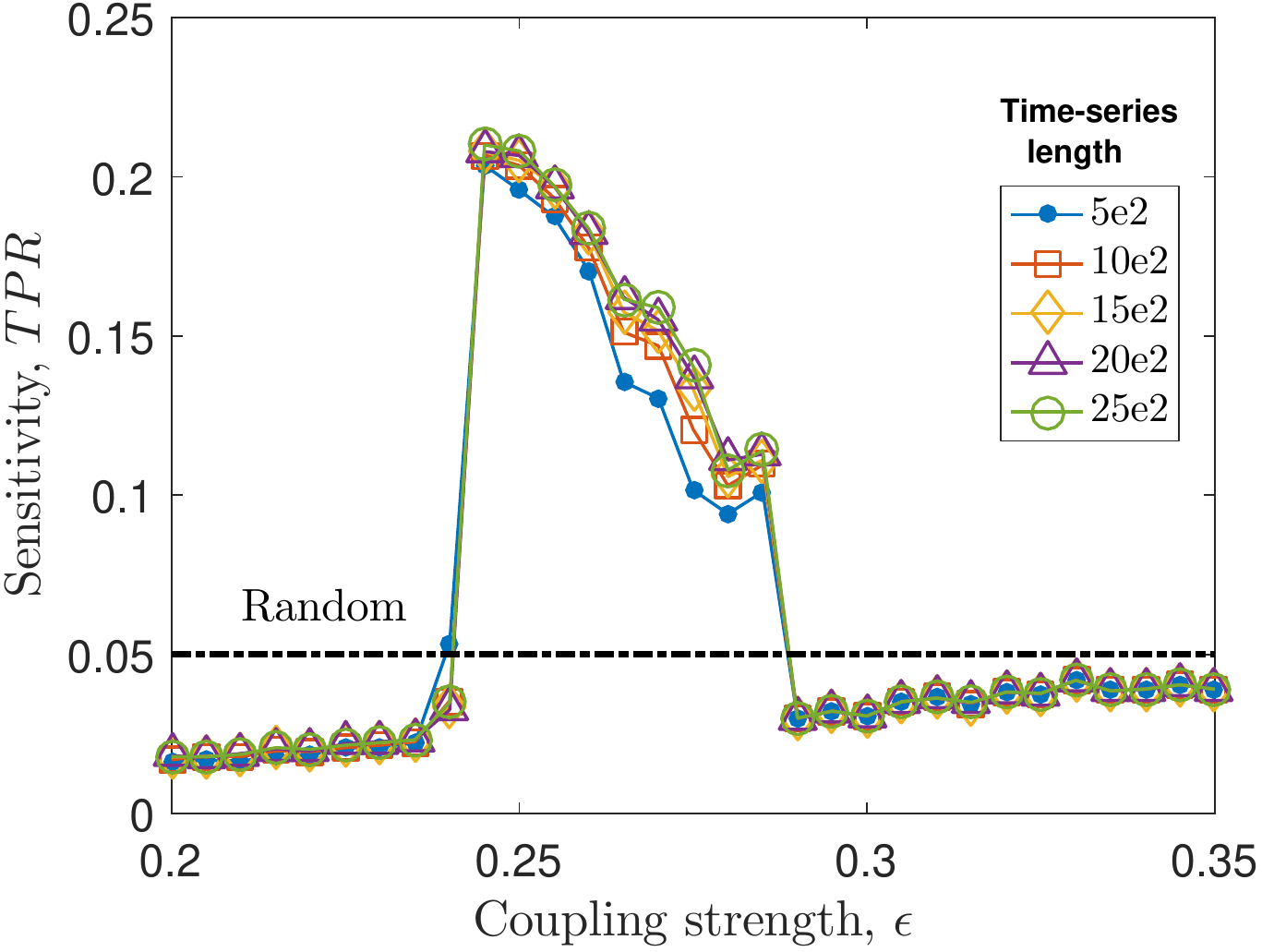}
   \end{center}
  \end{minipage}
 \end{center}
    \caption{{\bf Effect of the time-series length used on the efficiency when inferring Erd{\"o}s-R{\'e}nyi networks.} Panel \textbf{(a)} [\textbf{(b)}] shows the sensitivity, $TPR$, as a function of the coupling strength, $\epsilon$, when inferring $N=131$ [$N=277$] ER networks using membrane potentials with different time-series length. All points represent an average among $20$ different realisations and $10$ different initial conditions.} 
 \label{FIG:TPRCE_difftimelengths}
\end{figure}

In all our network inference processes we used a fixed time-series length ($T=5\times10^4$ when using membrane potentials and $T=2\times10^3$ when using interspike intervals). To check the robustness of our results against changes in the length of the time-series used to calculate cross-correlations between pairs of neurons, we calculate the same $TPR(\epsilon)$ curves shown in Fig.~\ref{FIG:TPR_vs_Ep_CE_WS_ER}, but using different time-series lengths. Fig.~\ref{FIG:TPRCE_difftimelengths} shows a comparison between the sensitivity as a function of the coupling strength for Erd{\"o}s-R{\'e}nyi networks, using different time-series lengths. We observe a great similarity between all the $TPR(\epsilon)$ curves, with only quantitative differences when the time-series is too short. We obtain similar results in the Watts-Strogatz and C. elegans case. Therefore, we can confirm that our findings are independent of the time-series length considered in our calculations.  

\subsection*{ROC Analysis}
 
 The analysis of the Receiver Operating Characteristic (ROC) curve is a potent tool for evaluating the efficiency of a classification algorithm \cite{brown2006receiver,fawcett2006introduction,rogers2005bayesian}. It has been used, for example, to assess the efficiency of clinical test for distinguishing between infected and healthy patients \cite{swetsevaluation}, or for testing machine-learning-based classification algorithms \cite{chawla2003smoteboost}. In the network inference scene, ROC analysis is often used to test and optimise inference methods \cite{rubido2014exact}. To assess the efficiency of an inference technique, ROC analysis relies on four fundamental quantities: the \textit{Sensitivity}, or \textit{True Positive Rate} ($TPR$, the fraction of present links that were correctly identified as such), \textit{Specificity}, or \textit{True Negative Rate} ($TNR$, the fraction of absent links that were correctly identified as such), \textit{False Positive Rate} ($FPR$, the fraction of absent links that were incorrectly identified as present), and the \textit{False Negative Rate} ($FNR$, the fraction of existing links that were incorrectly identified as absent).                                 

In the network inference context, these four quantities are naturally related by two constrains: the total number of links ($N[N-1]/2$ in a network with $N$ nodes) and the number $M$ of links conserved by the inference method. It can be readily shown that these constrains relate the four ROC analysis quantities according to Eq.~\ref{EQ:ROC_deffs}.
\begin{equation}
    \centering
    \begin{cases}
    TPR+FNR=1\\
    FPR+TNR=1.
    \end{cases}
    \label{EQ:ROC_deffs}
\end{equation}
\noindent 
 Hence, all the information that ROC analysis provides about the inference method is contained in two of the four fundamental variables. Furthermore, as we have shown in the Results section, if we have an estimate on the link density we can condense all the ROC analysis information in just one variable (e.g. the $TPR$).


\end{document}